\documentclass[%
 reprint,
superscriptaddress,
showpacs,
 amsmath,amssymb,
 aps,
longbibliography,
]{revtex4-1}

\pdfoutput=1

\usepackage{graphicx}
\usepackage{dcolumn}
\usepackage{bm}
\usepackage{hyperref}
\usepackage[mathlines]{lineno}
\usepackage{xfrac} 
\usepackage {array}  
\usepackage{color}

\newcommand{\sL}{\mathcal{L}_{2D}}

\newcommand{\sF}{\mathcal{F}}

\begin{document}

\preprint{APS/123-QED}

\title{Optimized Designs for Very Low Temperature Massive Calorimeters}

\author{Matt Pyle}
\email{mpyle1@berkeley.edu}
\affiliation{Physics Department, University of California Berkeley}%

\author{Enectali \surname{Figueroa-Feliciano}}
\affiliation{Physics Department, Massachusetts Institute of Technology}%


\author{Bernard Sadoulet}
\affiliation{Physics Department, University of California Berkeley}%

\date{\today}

\begin{abstract}
The baseline energy-resolution performance for the current generation of large-mass, low-temperature calorimeters (utilizing TES and NTD sensor technologies) is $>2$ orders of magnitude worse than theoretical predictions. A detailed study of several calorimetric detectors suggests that a mismatch between the sensor and signal bandwidths is the primary reason for suppressed sensitivity. With this understanding, we propose a detector design in which a thin-film Au pad is directly deposited onto a massive absorber that is then thermally linked to a separately fabricated TES chip via an Au wirebond, providing large electron-phonon coupling (i.e. high signal bandwidth), ease of fabrication, and cosmogenic background suppression.  Interestingly, this design strategy is fully compatible with the use of hygroscopic crystals (NaI) as absorbers. An 80-mm diameter Si light detector based upon these design principles, with potential use in both dark matter and neutrinoless double-beta decay, has an estimated baseline energy resolution of 0.35\,eV, 20$\times$ better than currently achievable.  A 1.75\,kg ZnMoO$_{4}$ large-mass calorimeter would have a 3.5\,eV baseline resolution, 1000$\times$ better than currently achieved with NTDs with an estimated position dependence $\frac{\Delta E}{E}$ of 6$\times$10$^{-4}$, near or below the variations found in absorber thermalization in ZnMoO$_{4}$~ and TeO$_{2}$. Such minimal position dependence is made possible by forcing the sensor bandwidth to be much smaller than the signal bandwidth. Further, intrinsic event timing resolution is estimated to be $\sim$170 $\mu$s for 3\,MeV recoils in the phonon detector, satisfying the event-rate requirements of large $Q_{\beta \beta}$ next-generation neutrinoless double-beta decay experiments. Quiescent bias power for both of these designs is found to be significantly larger than parasitic power loads achieved in the SPICA/SAFARI infrared bolometers.



\end{abstract}


\maketitle


\section{Motivation}

The success of experiments that use massive very low-temperature calorimeters (e.g. CRESST \cite{CRESSTII_2014_LowMass}, CUORE \cite{CUORICINO_PRC_08}, EDELWEISS \cite{EDELWEISS_11} and CDMS \cite{CDMSII_HE}) in rare event searches is quite natural since phonon vibrational modes with energy above $k_{b}$T freeze out and thus do not contribute to the crystal heat capacity, leading to $T^{3}$ heat capacity scaling for semi-conducting and insulating crystals. Consequently, detectors with a large active mass and excellent energy resolution should be possible. To reiterate the utility of this natural scaling law: for a given energy deposition, a giant 1-tonne Si crystal at 10 mK will have the same temperature change as 1 g at 1 K .

At low temperature ($\sim$100 mK), calorimetry based on Transition Edge Sensor (TES) technology is quite mature and has been used for energy measurement at virtually all energy scales from infrared-photon ($\sim$100 meV) \cite{NIR_TES_08} to alpha detection ($\sim$10 MeV) \cite{alphaTES_08}. The measured energy resolution of well designed devices throughout this regime roughly matches the theoretical expectation of
\begin{equation}
	\sigma_{E}^{2} = \frac{4 k_{b} T_{c}^{2} C}{\alpha} \sqrt{\frac{n \sF(T_{c},T_{b}) \xi(I)}{1-\frac{T_{b}^n}{T_{c}^n}}}	
\label{eq:Eres_lim}	
\end{equation}
where $T_{c}$ is the superconducting transition temperature of the TES, $\alpha$ ($\frac{T}{R}\frac{\partial R}{\partial T}$) is the unitless sensitivity parameter, and the unitless terms within the square root all combine to usually be of order 2 \cite{Irwin_TES}.  

\begin{figure}[h!]
\centering
\includegraphics[width=2.9in]{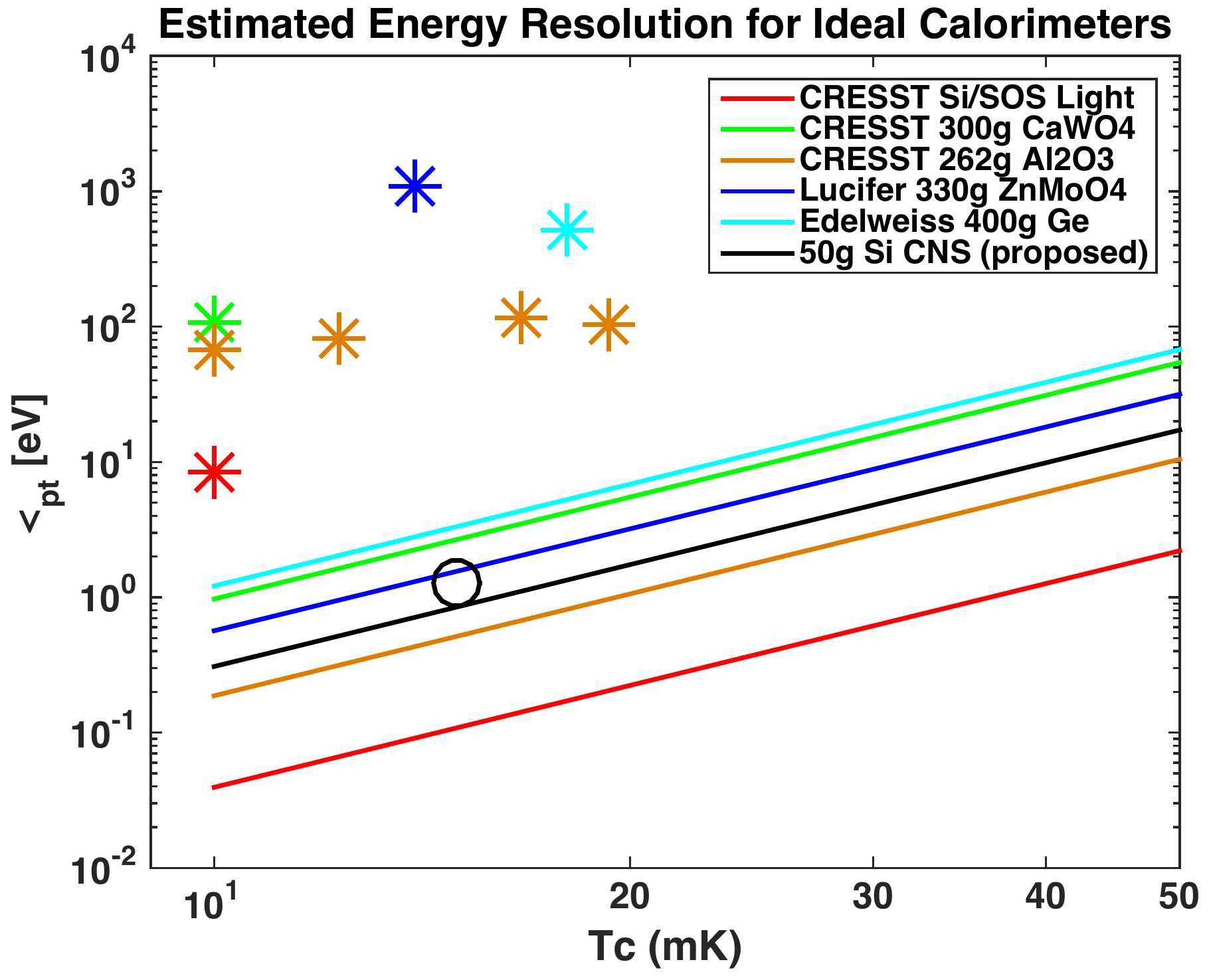}
\caption{\small{The naively estimated intrinsic energy resolution (Eq.~\ref{eq:Eres_lim}) as a function of $T$ for a variety of different massive calorimeters  versus their actual device performance (stars) \cite{CRESST_2009_Commissioning,CRESSTII_2014_LowMass,CRESST_2001_PhononDet,CUORE_ZnMoO4,EDELWEISS_11}. The plotted 50 g Si coherent neutrino scattering resolution (black circle) is simply a more detailed performance estimate by \cite{Tali_11_CNS}, taking into account internal device thermal conductances and other non-ideal device characteristics,  that shows rough agreement with the simpler scaling law estimates.}}
\label{fig:IrwinEres_T}
\end{figure}

In Fig. \ref{fig:IrwinEres_T}, we benchmark the measured performance of the current generation of massive calorimeters operating near 10 mK (stars) and a detailed proposal for a 50 g Si coherent neutrino scattering detector \cite{Tali_11_CNS} (black circle) against the theoretical scaling law of Eq.~\ref{eq:Eres_lim} assuming that the total heat capacity of the detector is 2$\times$ that of the absorber and that $\alpha = 20$ (values of 10--500 are common). Unfortunately we find that the energy-resolution performance of all current experiments is 2--3 orders of magnitude worse than expected from Eq.~\ref{eq:Eres_lim}. Consequently, either there are substantial experimental limitations at very low temperature that are not taken into account in Eq.~\ref{eq:Eres_lim}, and  \cite{Tali_11_CNS}, or the current generation of massive calorimeters could be significantly improved.


To gain insight into this energy resolution discrepancy, we will carefully study both the dynamics and noise of the original \cite{CRESST_2009_Commissioning} and composite \cite{CRESSTII_2014_LowMass} CRESST phonon detectors  as well as other calorimeters and then motivate and develop 6 design criteria that are applicable to very low-temperature massive detectors:
\begin{enumerate}
	\item The signal bandwidth (frequency scale at which energy transfers from the absorber to the sensor) must be greater than or equal to sensor bandwidth,
	\item The transition edge sensor (TES) must not phase separate,
	\item The sensor bandwidth must be larger than the $1/f$ noise threshold,
	\item The sensor bandwidth must be large enough to satisfy event rate requirements, and
	\item Microfabrication techniques should be used only on standard thin wafers.
	\item For applications that require minimal position dependence of energy estimators, we require that that signal bandwidth $\gg$ than the sensor bandwidth.
\end{enumerate}
Finally, we present 2 prototype designs that simultaneously achieve all of these design requirements:  a single-photon-sensitive light detector and a 1.75\,kg ZnMoO$_{4}$ double-beta decay detector.

\section{Simulating CRESST Phonon Detector}
\label{sec:DymNoise}

\begin{table} [h!]
\centering
\small
\begin{tabular}{| l | m{2.8cm} | m{2.4cm} | m{2.5cm} |}
\hline
\multicolumn{4}{| c |}{CaWO$_{4}$ absorber} \\
\hline
$V_{a}$	& Absorber volume						&														& $\pi$x2$^{2}$x4cm$^3$ \cite{Emilija_2008_thesis}\\ 
$M_{a}$	& Absorber mass						&														& 300 g \cite{Emilija_2008_thesis}\\ 
$C_{a}$	& Absorber heat capacity 					& $\Gamma_{\text{\tiny{CaWO$_{4}$}}} V_{a} T^{3}$					&132 $\mathrm{\frac{pJ}{K}}$@10mK\\ 
\hline
\multicolumn{4}{| c |}{W  TES} \\
\hline
$A_{t}$	& Cross sectional area					&														& 7.5mmx200nm\cite{Emilija_2008_thesis} \\ 
$l_{t}$	& Length								&														& 5.9 mm \cite{Emilija_2008_thesis} \\ 
$V_{t}$	& Volume								& $A_{t} l_{t}$												& 8.9x10$^{-3} $mm$^{3}$ \\ 
$C_{t}$	& Heat capacity 			 			& $f_{sc} \Gamma_{W} V_{t} T$								& 22.7 $\mathrm{\frac{pJ}{K}}$ @10mK \\ 
$P_{ta}$	& Power flow from TES to absorber			& $\Sigma_{epW} \! V_{t}\!(T_{t}^{n}\!-\!T_{a}^{n})$					& \\
$G_{ta}$	& Thermal conductance between TES and absorber %
											& $\frac{\partial P_{ta}}{\partial T_{t}}$
																									&142 $\mathrm{\frac{pW}{K}}$ @10mK\\ 
$G_{t\,int}$& internal TES conductance				& $\frac{L_{wf}}{\rho_{W}} \frac{A_{t}}{l_{t}/2} T$					&1.63 $\mathrm{\frac{nW}{K}}$\\ 
$R_{n}$	& TES normal resistance					& $\rho_{W} \frac{l_{t}}{A_{t}}$									& 300 $\mathrm{m\Omega}$ \cite{Emilija_2008_thesis} \\ 
$T_{c}$	& Transition temperature			&																& $\sim$ 10 mK \\ 
\hline
\multicolumn{4}{| c |}{Au wirebond: thermal link to bath} \\
\hline
$T_{b}$		& Bath temperature					&														& 6 mK \cite{CRESST_2005_FirstLimits} \\ 
$l_{tb}$		& $\sim$ Wirebond length	 			&														& 2 cm \cite{CRESST_2001_PhononDet} \\ 
$A_{tb}$		& Cross sectional area				&														& $\pi$ x 12.5$^{2} \,\mu$m$^{2}$ \cite{Emilija_2008_thesis}\\ 
$P_{tb}$		& Power flow from TES to bath			& $\frac{v_{f}\!d_{e}\!\Gamma\!A_{tb}}{6 l_{tb}}\!\left(T_{t}^{2}-T_{b}^{2}\right)$									& 19.2 pW\\
$G_{tb}$		& Thermal conductance from TES to bath %
								 			& $\frac{\partial P_{tb}}{\partial T_{t}}$
																									& 7.5 nW/K @10mK\\ 
\hline
\multicolumn{4}{| c |}{Electronics properties} \\
\hline
$L$			& Inductance (SQUID+ parasitic) 		&														& 350 nH \cite{CRESST_2007_electronics} \\ 
$R_{l}$		& Load resistor (shunt+ parasitic)		&														& 40 m$\Omega$ \cite{CRESST_2007_electronics}\\ 
$T_{l}$		& Temperature of $R_{l}$ 				&														& 10mK \cite{CRESST_2007_electronics} \\
$S_{I SQUID}$& SQUID Current Noise 				&														& 1.2 pA/$\sqrt{\text{hz}}$\cite{CRESST_2007_electronics}\\ 
\hline
\end{tabular}
\caption{\small{Estimated CRESST-II phonon-detector device parameters using the material properties in Appendix \ref{appendix:MatProp}.}}
\label{tab:PhononDet}
\end{table}

\begin{figure}[ht]
\centering
\includegraphics[width=3.3in]{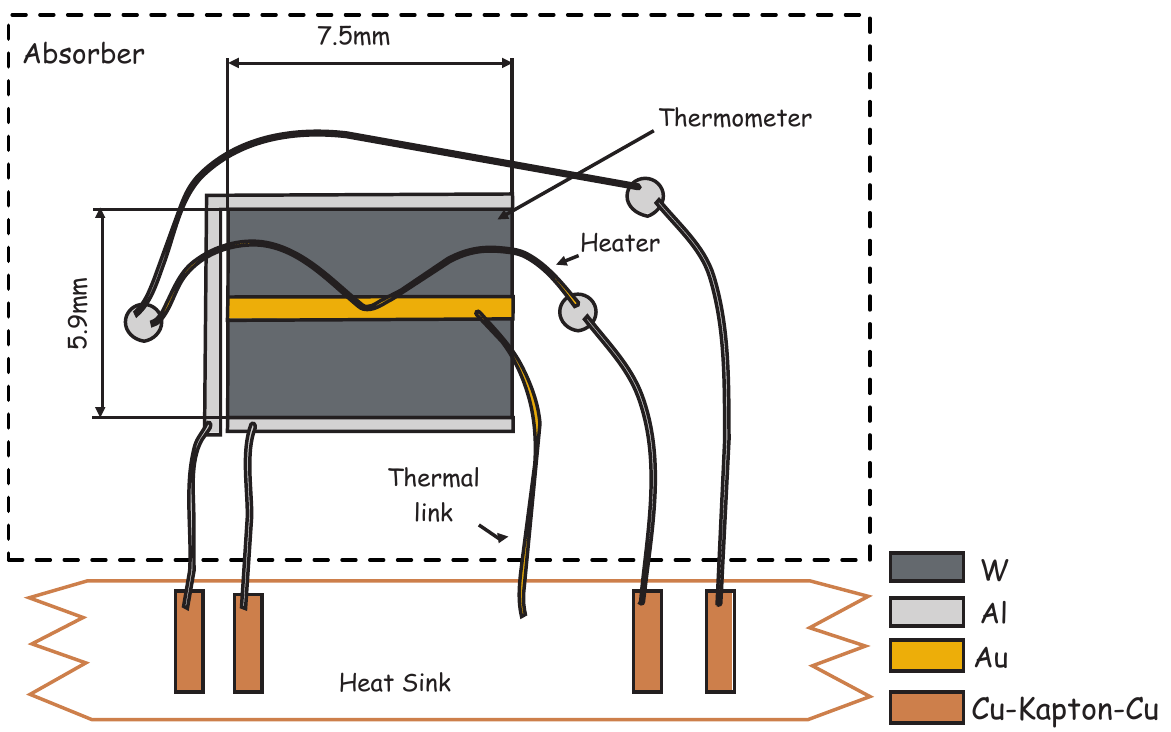}
\caption{\small{Schematic of the TES and its connections on the original CRESST-II phonon detector \cite{Emilija_2008_thesis}.}}
\label{fig:CaWO4_Qdiag}
\end{figure}

Modeling of complex calorimeters with multiple coupled thermal degrees of freedom (DOF) like the CRESST phonon detector has been a very active area of research, particularly within the x-ray and gamma-ray TES communities \cite{Tali_Thesis,TaliComplexuCal,Smith_Hydra,Irwin_GammaTES}. Thus, 
we need only apply mature strategies, being careful to follow the conventions of \cite{Irwin_TES}. This is made even easier by the fact that CRESST has already modeled the dynamics of their detectors \cite{CRESST_1995_ProbstModel, CRESSTII_2009_SplitDetector, Emilija_2008_thesis} and even attempted to simulate noise \cite{Emilija_2008_thesis}; so we will concentrate in particular on developing analytical simplifications and  physical intuition, neither of which has yet been done.


\subsection{Dynamics}
The original (non-composite) CRESST-II phonon detector is a large 300 g cylindrical CaWO$_{4}$ crystal with a W TES (T$_{c} \approx$ 10 mK) and all of its required accessories (bonding pads, Al bias rails) fabricated directly upon its surface, as shown in Fig. \ref{fig:CaWO4_Qdiag} \cite{Emilija_2008_thesis}. Unfortunately, even after such painstaking fabrication efforts, the thermal coupling between the TES and the large mass absorber, $G_{ta}$, is still 50$\times$ smaller than the thermal conductance between the TES and the bath via the direct electronic coupling of the Au wirebond, $G_{tb}$ ({\textit cf.} Table. \ref{tab:PhononDet}). This is because the W electronic system (sensor) and the phonon system of the TES and absorber almost completely decouple at very low temperatures since the thermal power flow scales as $T^{5}$. 

\begin{figure}[htp!]
\centering
\includegraphics[width=2.0in]{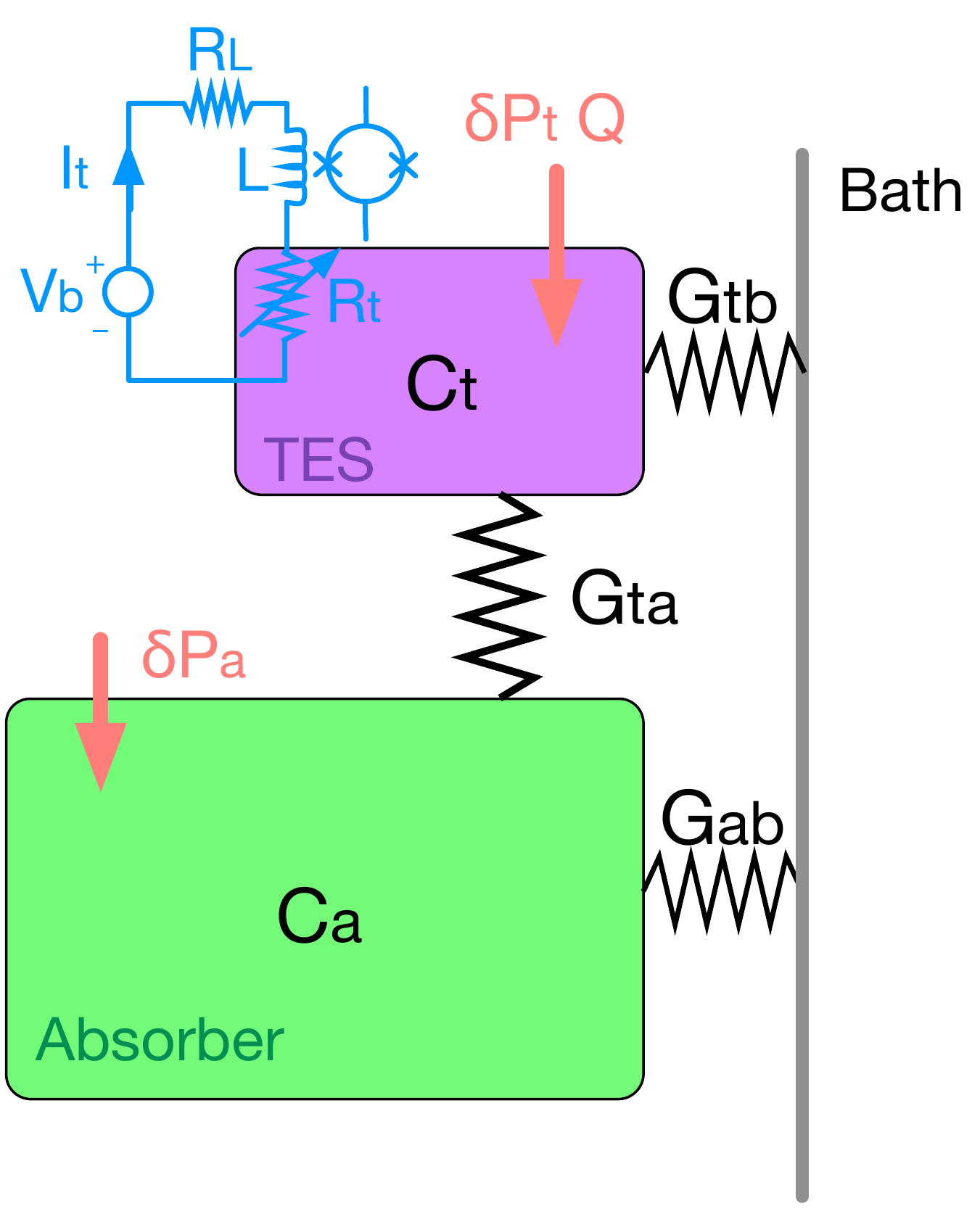}
\caption{\small{ Simplified model of the CRESST phonon and light detectors with 2 thermal DOF as well as the voltage-bias and current readout circuit}}
\label{fig:Det_2DModel}
\end{figure}

From a modeling perspective, the primary consequence is that any realistic device simulation requires separate thermal DOF for the absorber, $T_{a}$, and the TES, $T_{t}$, in addition to modeling the current flowing through the TES $I_{t}$ which will be measured through an inductively coupled SQUID. Thus in a way similar to CRESST \cite{CRESST_1995_ProbstModel}, we will model the system as 3 coupled non-linear differential equations  

\begin{equation}
\begin{split}
L \frac{dI_{t}}{dt}           &= V_{b} - I_{t} \left( R_{l}+R_{t}(T_{t},I_{t}) \right) + \delta V\\
C_{t}  \frac{dT_{t}}{dt}  &= I_{t}^{2}R_{t}(T_{t},I_{t}) - P_{tb}(T_{t},T_{b})- P_{ta}(T_{t},T_{a})  + Q + \delta P_{t}\\
C_{a} \frac{dT_{a}}{dt} &= - P_{ab}(T_{a},T_{b}) + P_{ta}(T_{t},T_{a}) + \delta P_{a}
\end{split}
\label{eq:NonLin_dyn}
\end{equation}
where $\delta V$ is a small voltage-bias excitation on top of the DC voltage-bias $V_{b}$ that we will use to model Johnson noise as well as the small signal dynamics of the detector. Likewise, $\delta P_{t}$ and $\delta P_{a}$ are power excitations into the TES and absorber respectively. This model can also be seen diagrammatically in Fig. \ref{fig:Det_2DModel}. Variable definitions and estimated sizes can be found in Table \ref{tab:PhononDet}. One relatively unique feature of the CRESST design is the capability to directly heat the TES electronic system through an additional heater circuit ($Q$). When held constant, this is effectively equivalent to being able to easily vary the bath temperature $T_b$ on a detector by detector basis.

Taylor expanding to first order around the operating point, $[I_{to},T_{to},T_{ao}]$ and then Fourier transforming, these equations simplify to 
\begin{widetext}
\begin{equation}
			\left[ \!
			\begin{array}{c c c}
				 j\omega + \frac{R_{to} (1+\beta) +R_{l}}{L}	& \frac{I_{to} R_{to} \alpha}{L T_{to} }                                    				& 0 \\
				 -\frac{I_{to} R_{to}(2+\beta)}{C_{t}}			& j \omega +\frac{\frac{I_{to}^{2}R_{to} \alpha}{T_{to}}- G_{ta}-G_{tb}}{C_{t}}	&   -\frac{G_{ta,a}}{C_{t}} \\
				 0                                                                       	&  -\frac{G_{ta}}{C_{a}} 		     				     					&   j \omega + \frac{G_{ta,a}+ G_{ab}}{C_{a}}
			\end{array}
			\! \right]
			\left[\!
			\begin{array}{c}
				\Delta I _{t}(\omega)\\
				\Delta T_{t}(\omega) \\
				\Delta T_{a}(\omega)
			\end{array}
			\!\right]
			=
			\left[\!
			\begin{array}{c}
				\frac{\delta V(\omega)}{L} \\
				\frac{\delta P_{t}(\omega)}{C_{t}} \\
				\frac{\delta P_{a}(\omega)}{C_a}
			\end{array}
			\!\right]				
\label{eq:3D_ODE_Fourier}
\end{equation}
which in the pertinent limit of  $L \rightarrow 0$ can be simplified to only thermal DOF 
\begin{equation}
		        \left[ \!
			\begin{array}{c c}
				   j\omega + (\sL\!+\!1) \frac{G_{ta} +G_{tb}}{C_{t}}	&   \frac{-G_{ta,a}}{C_{t}} \\
				   \frac{-G_{ta}}{C_{a}} 		     				     				&j \omega  + \frac{G_{ta,a}\! + \! G_{ab}}{C_{a}}
			\end{array}
			\! \right]
			\left[\!
			\begin{array}{c}
				\Delta T_{t} \\
				\Delta T_{a}
			\end{array}
			\!\right]
		=	
			\left[\!
			\begin{array}{c}
				\frac{\delta P_{t}}{C_{t}} + \frac{ I_{to}R_{to} (2+\beta) \delta V  }{C_{t} \left(R_{l} + R_{to}\left[1+\beta \right] \right)} \\
				\frac{\delta P_{a}}{C_a}
			\end{array}
			\!\right]				
\label{eq:2D_ODE}
\end{equation}
\end{widetext}
where 
\begin{equation}
\delta I_{t} = \frac{-I_{to} R_{to} \alpha}{\left( R_{l} + R_{to}\left[1+\beta\right]\right) T_{to}} \delta T_{t} + \frac{1}{R_{l} + R_{o}\left[1+\beta\right]} \delta V
\label{eq;dI_2D}
\end{equation}
As is standard, $\beta$ or ${\frac{I_{to}}{R_{to}} \frac{\partial R_{t}}{\partial I_{t}}(T_{to},I_{to}) |_{T_{t}}}$ characterizes the unwanted dependence of $R_{t}$ on the current. Also note that we have generalized Irwin's loop gain parameter $\mathcal{L}_{I}$ to
\begin{equation}
\sL = \alpha \frac{R_{to}-R_{l}}{R_{to}(1+\beta)+R_{l}}  \frac{I_{to}^{2}R_{to}}{\left(G_{tb}+G_{ta}\right)T_{to}}
\label{eq:LoopGain}
\end{equation}
to account for the 2 thermal DOF as well as the fact that CRESST has purposely chosen to use a load resistor ($R_{l}$) that is of similar magnitude to $R_{to}$ to suppress electro-thermal feedback. Finally, due to the possibility that $T_{to}$ and $T_{ao}$ could be macroscopically different, the thermal conductances $\frac{\partial P_{ta}}{\partial T_{t}}$ and $\frac{\partial P_{ta}}{\partial T_{a}}$ could have  different values and consequently we define 
\begin{equation}
\begin{split}
G_{ta,a} &= \frac{\partial P_{ta}}{\partial T_{a}}\\
	      &= \frac{\partial P_{ta}}{\partial T_{t}} \left(\frac{T_{a}}{T_{t}}\right)^{n_{ep}-1}\\
	      &= G_{ta} \left(\frac{T_{a}}{T_{t}}\right)^{n_{ep}-1}
\end{split}
\end{equation}
Inversion of the generalized impedance matrix, $M_{2D}$, found in Eq.~\ref{eq:2D_ODE} leads to the current transfer functions for the thermal power response for direct heating into the TES
\begin{equation}
\begin{split}
\frac{\partial I_{t}}{\partial P_{t}} 	&=	\frac{ - I_{to} R_{to}}{R_{l} \! + \! R_{to}(1 \! + \! \beta)} 
								\frac{\alpha}{C_{t}T_{t}} 
								M_{2D (1,1)}^{-1}(\omega) \\
							&\sim 		\frac{- I_{to} R_{to}}{R_{l} \! + \! R_{so}(1 \! + \!\beta)}
											\frac{\alpha}{\left( \sL\!+\! 1 \right)}
											\frac{1}{\left( G_{tb} \! + \! G_{ta} \right)T_{t}} \frac{1}{1+j\omega/\omega_\mathit{eff}}\\
							&\sim 	\frac{-1}{I_{to}(R_{to} \! - \! R_{l})} \frac{\sL}{\sL \! + \!1 } \frac{1}{1+j\omega/\omega_\mathit{eff}}
\end{split}
\label{eq:dIdPt}
\end{equation}
and into the absorber
\begin{equation}
\begin{split}			
\frac{\partial I_{t}}{\partial P_{a}} 	&=      		\frac{- I_{to} R_{to}}{R_{l} \! + \! R_{to}(1 \! + \! \beta)}
											\frac{\alpha}{C_{a}T_{t}}
											M_{2D (1,2)}^{-1}(\omega) \\
							&\sim	 \frac{\partial I_{t}}{\partial P_{t}}
									\left(\frac{G_{ta,a}}{G_{ta,a} \!+ \!G_{ab}}\right)
									\left(\frac{1}{1+j\omega/\omega_{ta}}\right) 												
\end{split}
\label{eq:dIdPa}
\end{equation}
where 
\begin{equation}
\begin{split}
	\omega_\mathit{eff} \! &=	\left(\sL \! + \! 1 \right) \frac{G_{ta} \! + \! G_{tb}}{C_{t}} \! + \!  
					\frac{1}{\sL \! + \! 1} \frac{G_{ta,a}}{G_{ta} \! + \! G_{tb}} \frac{G_{ta}}{C_{a}} \! + \! \dots\\
	\omega_{ta} \! &=	\frac{G_{tb} \! + \! G_{ta,a}}{C_{a}}  \! -\! 
					\frac{1}{\sL \! + \! 1} \frac{G_{ta,a}}{G_{ta}+G_{tb}} \frac{G_{ta}}{C_{a}} \! + \! \dots
\end{split}
\label{eq:omega}
\end{equation}
All of the approximations shown are valid only for the pertinent limiting case where $G_{ta}$ and $G_{ab} \ll G_{tb}$ ({\textit cf.} Table \ref{tab:PhononDet}).  

Qualitatively, the last term of Eq.~\ref{eq:dIdPa}, which is an additional pole that surpresses the bandwidth of $\frac{\partial I}{\partial P_{a}}$ relative to $\frac{\partial I}{\partial P_{t}}$, is due to the fact that thermal energy must be transported across the tiny $G_{ta}$ before it is seen by the TES. Likewise, the middle term of Eq.~\ref{eq:dIdPa} is a constant suppression factor because a portion of $\delta P_{a}$ is shunted through $G_{ab}$. This suppression should be small for the CRESST phonon detector.  However, for the CRESST light detector, the Au wirebond pad on the substrate has an electron-phonon coupling that is larger than that of the W TES, and consequently this factor is significant \cite{Emilija_2008_thesis}. Another detector with significant shunting is found in  CUORE  where the  vast majority of the absorber thermal signal flows through $G_{ab}$ \cite{CUORE_ThermalConductance_92}.

Both the $\frac{\partial I}{\partial P_{t}}$ (Eq.~\ref{eq:dIdPt}) and $\frac{\partial I}{\partial P_{a}}$ (Eq.~\ref{eq:dIdPa}) transfer functions are important for understanding the signal response of massive calorimeters, because some of the high-energy athermal phonons produced by a particle recoil in the absorber may be collected and thermalized within the TES ($\delta P_{t}$) before they thermalize within the absorber  ($\delta P_{a}$), since the electron-phonon coupling varies so significantly with phonon energy and temperature. Consequently, a true particle recoil within the absorber should be modeled within our 2 thermal DOF system as 
\begin{equation}
\begin{split}
\frac{\partial I_{t}}{\partial E_{\gamma}} 	&= \frac{\partial I_{t}}{\partial P_{t}} \frac{\partial P_{t}}{\partial E_{\gamma}} + \frac{\partial I_{t}}{\partial P_{a}} \frac{\partial P_{a}}{\partial E_{\gamma}}\\
								&= \frac{\partial I_{t}}{\partial P_{t}} \left( \frac{\partial P_{t}}{\partial E_{\gamma}} + \frac{G_{ta,a}}{G_{ta,a} \!+ \!G_{ab}}
									\frac{1}{1+j\omega/\omega_{ta}} \frac{\partial P_{a}}{\partial E_{\gamma}}\right)
\end{split}
\label{eq:dItdE}
\end{equation} 
where the benefit of being able to write $\frac{\partial I}{\partial P_{a}}$ in terms of $\frac{\partial I}{\partial P_{t}}$ plus additional factors is readily apparent. 

\subsection{Noise Estimation}

With the dynamical response of the detector now modeled, we can estimate the magnitude of noise from thermal power fluctuations across $G_{tb}$, $G_{ab}$, and $G_{ta}$, the Johnson noise across $R_{t}$ and $R_{l}$ and finally the first stage squid noise and compare their relative sizes by referencing them to a thermal power signal flowing directly into the TES ($\delta P_{t}$).  This reference point was chosen purposely so that intuition from simpler 1 DOF thermal systems could be used and to suppress differences due to CRESST's use of a non-standard electronics readout scheme with a large $R_{l}$ to suppress electro-thermal feedback.

With this choice of reference, thermal fluctuation noise (TFN) across $G_{tb}$ is flat and can be estimated as
\begin{equation}
S_{P_{t} \, G_{tb}}(\omega) = 4k_{b}T_{to}^{2}G_{tb} \sF(T_{to},T_{b},\mbox{diffusive}) 
\label{eq:Spt-Gtb}
\end{equation}
where  $\sF$ is a noise suppression term with a value between \sfrac{1}{2} and 1 to account for the fact that the power noise across a thermal link between two different temperatures (a non-equilibrium situation) is less than the naively derived equilibrium noise  \cite{McCammon_TES}.
%
Now, estimation of the TFN across $G_{ab}$ is slightly more difficult since we must refer it to the TES and thus
\begin{equation}
\begin{split}
S_{P_{t} \, G_{ab}} 	&= \left({\frac{\partial I_{t}}{\partial P_{a}}}/\frac{\partial I_{t}}{\partial P_{t}}\right)^2 4k_{b}T_{ao}^{2}G_{ab} \sF(T_{ao},T_{b},\mbox{diffusive})  \\
				&\sim {\left( \frac{G_{ta,a}}{G_{ta,a}\! + \!G_{ab}}\right)}^{2}
				\frac{1}{1+\omega^{2}/\omega_{ta}^2}
				4k_{b}T_{ao}^{2}G_{ab} \sF
\end{split}
\label{eq:Si-Gab}
\end{equation}
This noise is subdominant, even below the $\omega_{ta}$ pole since $G_{ab} \ll G_{tb}$ for both CRESST detectors (as well as for our new design).

Next, we estimate our sensitivity to thermal fluctuations between the electronic and phonon systems (i.e. across $G_{ta}$). We must be cognizant that these fluctuations are anti-correlated to  conserve energy;  if thermal power randomly flows into the phonon system, it must be flowing out of the electronic system and vice-versa. Thus, our sensitivity is
%
\begin{equation}
\begin{split}
S_{P_{t} \, G_{ta}} 	&= \left( \frac{\frac{\partial I_{t}}{\partial P_{t}} - \frac{\partial I_{t}}{\partial P_{a}}}{\frac{\partial I_{t}}{\partial P_{t}} } \right)^{2}   4k_{b}T_{to}^{2}G_{ta} \sF(T_{to},T_{ao},\mbox{ballistic})\\
				&\sim \left( 1- \frac{G_{ta,a}}{G_{ta,a}+G_{ab}} \frac{1}{1+j \omega/\omega_{ta}}\right)^{2} 4k_{b}T_{to}^{2}G_{ta} \sF
\end{split}
\label{eq:Spt-Gta}
\end{equation}
The insensitivity to this noise below $\omega_{ta}$ is a direct ramification of energy conservation and is a general feature of all ``massless'' internal thermal conductances. Of course, for the CRESST designs, this noise is again negligible compared to that from $G_{tb}$, simply due to their relative sizes.

TES Johnson noise is by far the most challenging to calculate for 3 reasons.  One must take into account the anti-correlation between $\delta V$ and $\delta P_{t}$ \cite{Irwin_TES,McCammon_TES} as well as the noise boost due to current sensitivity (non-zero $\beta$) \cite{Irwin_TES}. Finally, we must reference back to the TES input power.  We do this in 2 steps. First, we calculate the Johnson noise referenced to current flowing through the TES ($\delta_{It}$) 
%
\begin{equation}
\begin{split}
S_{I_{t} \, R_{t}} 	&= \frac{4 k_{b} T_{to} R_{to} (1 + \beta)^{2}}{(R_{l} \! + \! R_{o}\left[1\! + \! \beta\right])^{2}} \! 
				\left( 1 + \frac{\partial \! I_{t}}{\partial \! P_{t}} \! I_{to}(R_{to}\! - \!R_{l}) \right)^{2}\\
			         &\sim   \frac{4 k_{b} T_{to} R_{to} (1 + \beta)^{2}}{(R_{l} \! + \! R_{o}\left[1\! + \! \beta\right])^{2}}
			         \left( 1 - \frac{\sL}{\sL+1} \frac{1}{1+j\omega/\omega_\mathit{eff}} \right)^{2}\\        
\end{split}
\label{eq:Spt-Rt}
\end{equation}
where we see that the anti-correlation leads to noise suppression at low frequencies for high $\sL$. Referencing to $\delta P_{t}$, we obtain:
\begin{equation}
\begin{split}
S_{P_{t} \, R_{t}}  \! 	&= \frac{4 k_{b} T_{to} R_{to} (1 + \beta)^{2}}{(R_{l} \! + \! R_{o}\left[1\! + \! \beta\right])^{2}} \! 
				\left( \frac{1}{\frac{\partial \! I_{t}}{\partial \! P_{t}}}+\! I_{to}(R_{to}\! - \!R_{l}) \right)^{2}\\
			   \!   &\sim 4 k_{b} T_{to}^2 (G_{ta} \! + \! G_{tb}) \frac{ (1 \! + \! \beta)^{2}}{\alpha^{2} \frac{I_{to}^{2}R_{to}}{(G_{ta}+G_{tb})T_{to}}}\!
			   {\left( 1 \! + \! \frac{j \omega}{\frac{\omega_\mathit{eff}}{\sL+1}}\right)}^{2}
			   %
\end{split}
\label{eq:Sp_Rt}
\end{equation}
Again, we find that as long as $\alpha$ is large and heating of the TES via external heating ($Q$) or a bath temperature ($T_{b}$) near $T_{c}$ does not suppress $I_{to}^{2} R_{to}$ too significantly, then the thermal fluctuations across $G_{tb}$ dominate the noise at low frequency. However, this is clearly not true at high frequencies due to the zero at $\frac{\omega_\mathit{eff}}{\sL+1}$.  In fact, it is this TES Johnson noise term that sets the bandwidth for optimal energy estimators of $\delta P_{t}$, which we can estimate by finding the frequency beyond which  $S_{P_{t}\,R_{t}} > S_{P_{t}\,G_{tb}}$. This frequency, which we will designate as the signal-to-noise bandwidth, is given by:
%
\begin{equation}
\omega_{S/N \, \delta \! P_{t}} \sim \frac{G_{ta}+G_{tb}}{C_{t}} \frac{\alpha}{1+\beta} \sqrt{\sF \frac{G_{tb}}{G_{tb}+G_{ta}}  \frac{I_{to}^{2}R_{to}}{(G_{ta}+G_{tb})T_{to}} }
\label{eq:w_OF_Pt}
\end{equation}

\begin{figure}[h!]
\centering
\includegraphics[width=3.3in]{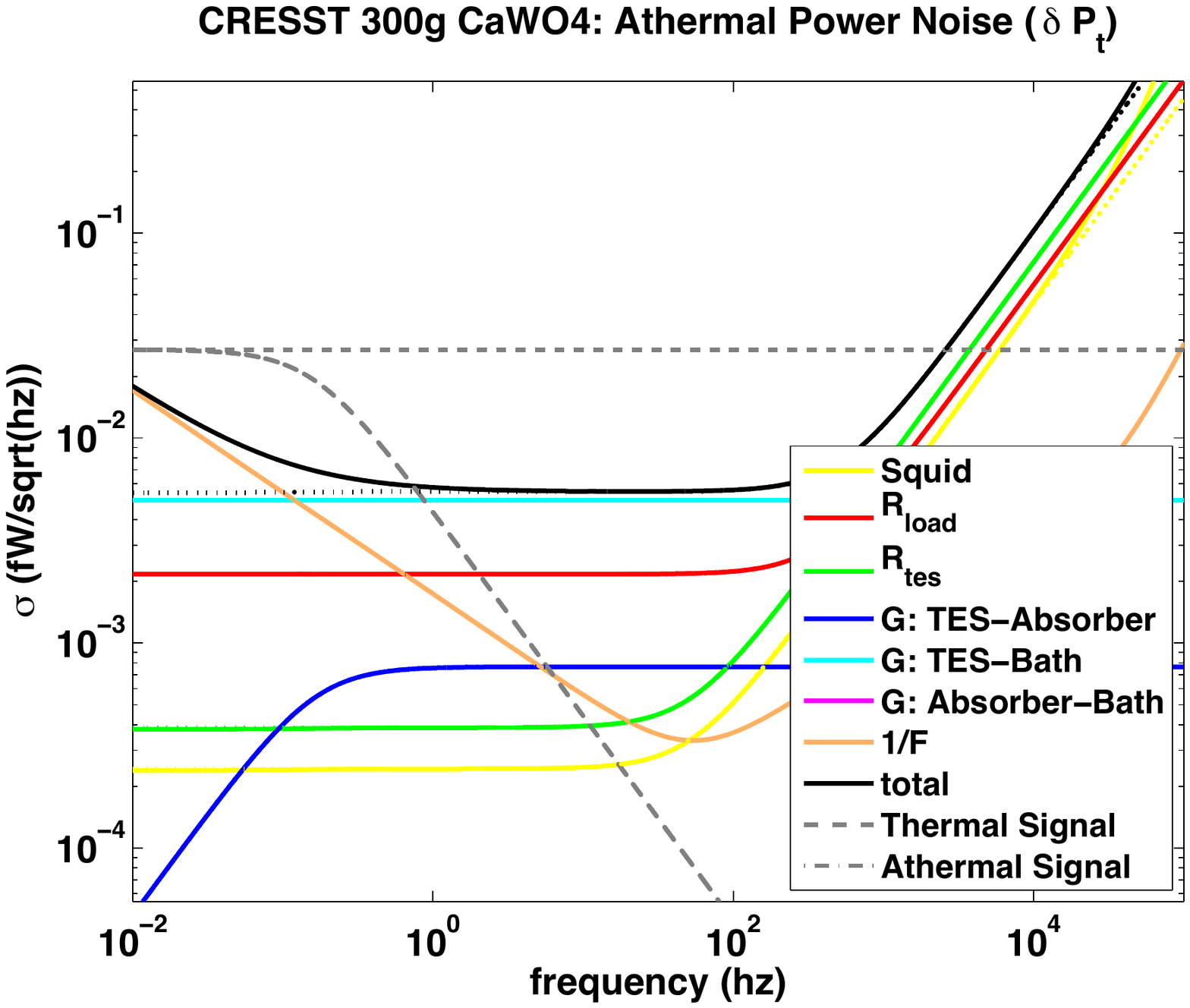}
\includegraphics[width=3.3in]{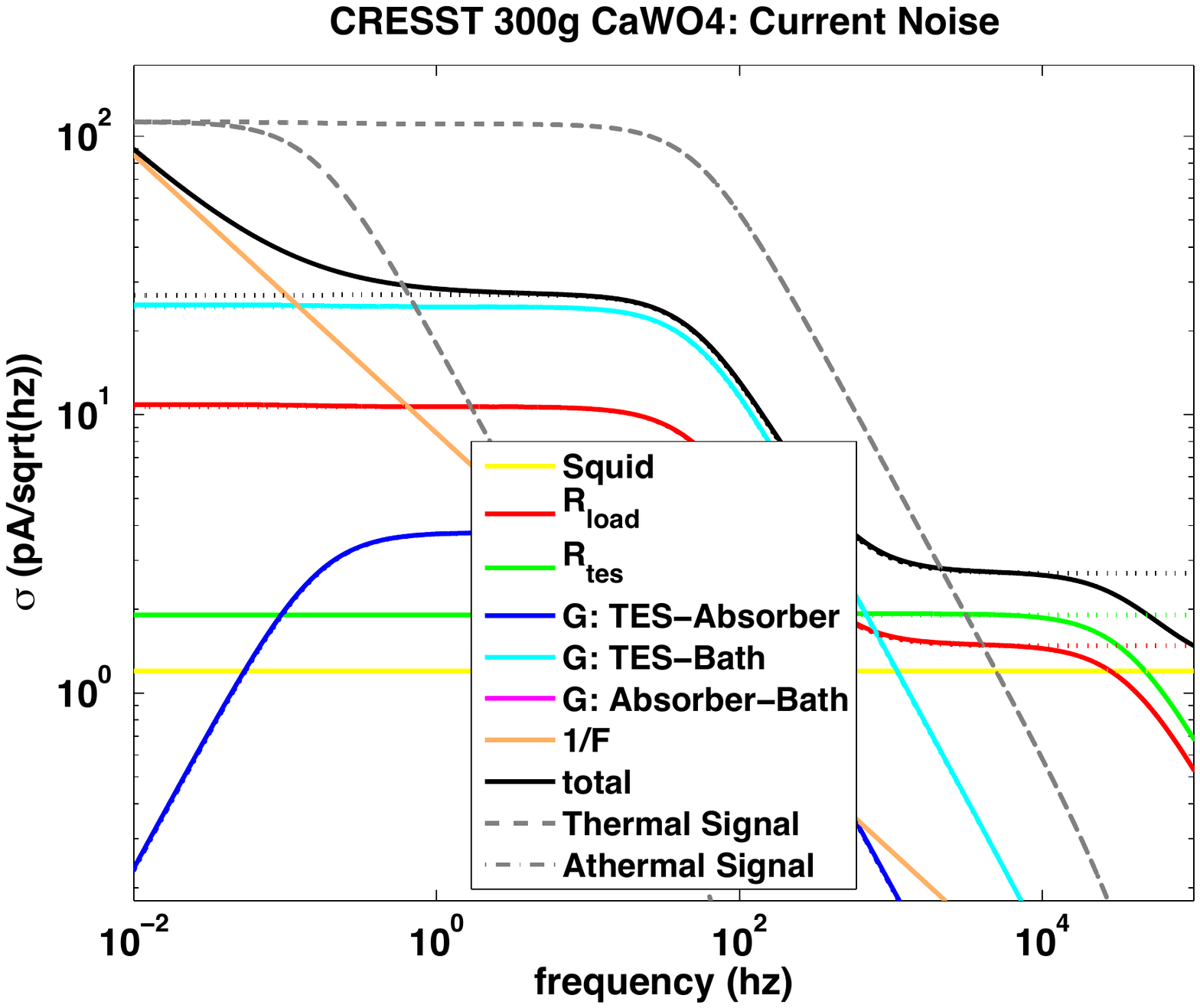}
\caption{\small{Simulated noise spectrum for CRESST phonon detector referenced to TES power $\delta P_{t}$(top) and TES current $\delta I_{t}$ (bottom). Dotted lines correspond to Taylor expansions in the limit of $L\rightarrow$0 and $G_{ta}$ \& $G_{ab}\ll G_{tb}$ and no $1/f$ noise.}}
\label{fig:Spt-Sit}
\end{figure}

For electronics designs with large electro-thermal feedback, $\omega_{S/N \, \delta P_{t}} \sim \omega_\mathit{eff}$.  However, for CRESST where $\sL$ has purposely been set to near 0 by choosing $R_{l} \sim R_{to}$, $\omega_{S/N \, \delta \! P_{t}} \gg \omega_\mathit{eff}$. 
To reiterate, when referencing both the signal and noise to current (Fig. \ref{fig:Spt-Sit} bottom), both the current signal and the total current noise have a pole at $\omega_\mathit{eff}$, and thus the signal-to-noise (which is equivalent to rereferencing to TES power) remains flat all the way up to $\omega_{S/N \, \delta \! P_{t}}$.

Unfortunately, neither the squid noise
%
\begin{equation}
S_{P_{t} SQUID}=\frac{1}{\frac{\partial I_{t}}{\partial P_{t}}} S_{I_{t} SQUID}
\label{eq:Spt-squid}
\end{equation}
nor the Johnson noise from the load resistor, R$_{l}$,
%
\begin{equation}
S_{P_{t} \, R_{l}}  \! = \frac{4 k_{b} T_{to} R_{to} (1 + \beta)^{2}}{(R_{l} \! + \! R_{o}\left[1\! + \! \beta\right])^{2}} \! 
				\left( \frac{1}{\frac{\partial \! I_{t}}{\partial \! P_{t}}}+\! I_{to}R_{to}(2 \! + \! \beta) \right)^{2}
\label{eq:Spt-Rl}
\end{equation}
can be trivially written in terms of $G_{tb}$.  However, as shown in Fig. \ref{fig:Spt-Sit} they are subdominant to the combination of $G_{tb}$ TFN and $R_{t}$ Johnson noise (this latter can also be seen through comparison of Eq.~\ref{eq:Spt-Rl} to Eq.~\ref{eq:Spt-Rt}).

Finally, we would ideally like to have an estimate of the $1/f$ noise found in the current CRESST experimental setup, particularly since there are some indications that it could be an important contributor to the overall noise in CRESST currently \cite{Emilija_2008_thesis}. More importantly, we would like to estimate the 1/$f$ noise expected in a next-generation experimental setup to assess its effect on the device design. For the former, significant effort would be required that is beyond the scope of this paper.  For the latter, we will use the noise performance of the  SPIDER TES bolometers, which begin to be dominated by $1/f$ at 0.1 Hz, as guide to what could be achieved with effort, since CMB experiments are quite sensitive to low frequency noise \cite{SPIDER} and thus we will set a design goal of $\omega_{S/N \, \delta \! P_{t}} > $ 0.5 hz. 
 
\section{TES Phase Separation}
\label{sec:PhaseSep}

Unfortunately, as first realized in the earliest days of CRESST calorimeter development \cite{CRESST_1995_ProbstModel}, the simple 2 DOF block thermal model shown in Fig. \ref{fig:CaWO4_Qdiag} that we have used for our CRESST-II phonon-detector resolution estimates is not entirely accurate, since  $G_{tb}$ is also significantly larger than the internal thermal conductance of the W TES, $G_{t\,int}$({\textit cf.} Table. \ref{tab:PhononDet}). In this situation, the difference in temperature between the TES $T_{c}$ and the effective bath temperature is mostly internal to the TES, and thus there is a significant temperature gradient across the sensor. In devices with large $\alpha$, this thermal gradient is further exasperated by positive-feedback effects of joule heating and the TES can separate into sharply defined superconducting and normal regions, with only a very thin region within the transition that is sensitive to temperature fluctuations \cite{blas_overview}.  

Although a computational simulation of phase separation within the CRESST geometry is beyond the scope of this paper (and of dubious value without a very careful matching to experimental data since the dynamics depend on the resistivity of the entire transition, rather than at a single operation point), we can qualitatively discuss the ramifications.

\begin{figure}[h!]
\centering
\includegraphics[width=3.3in]{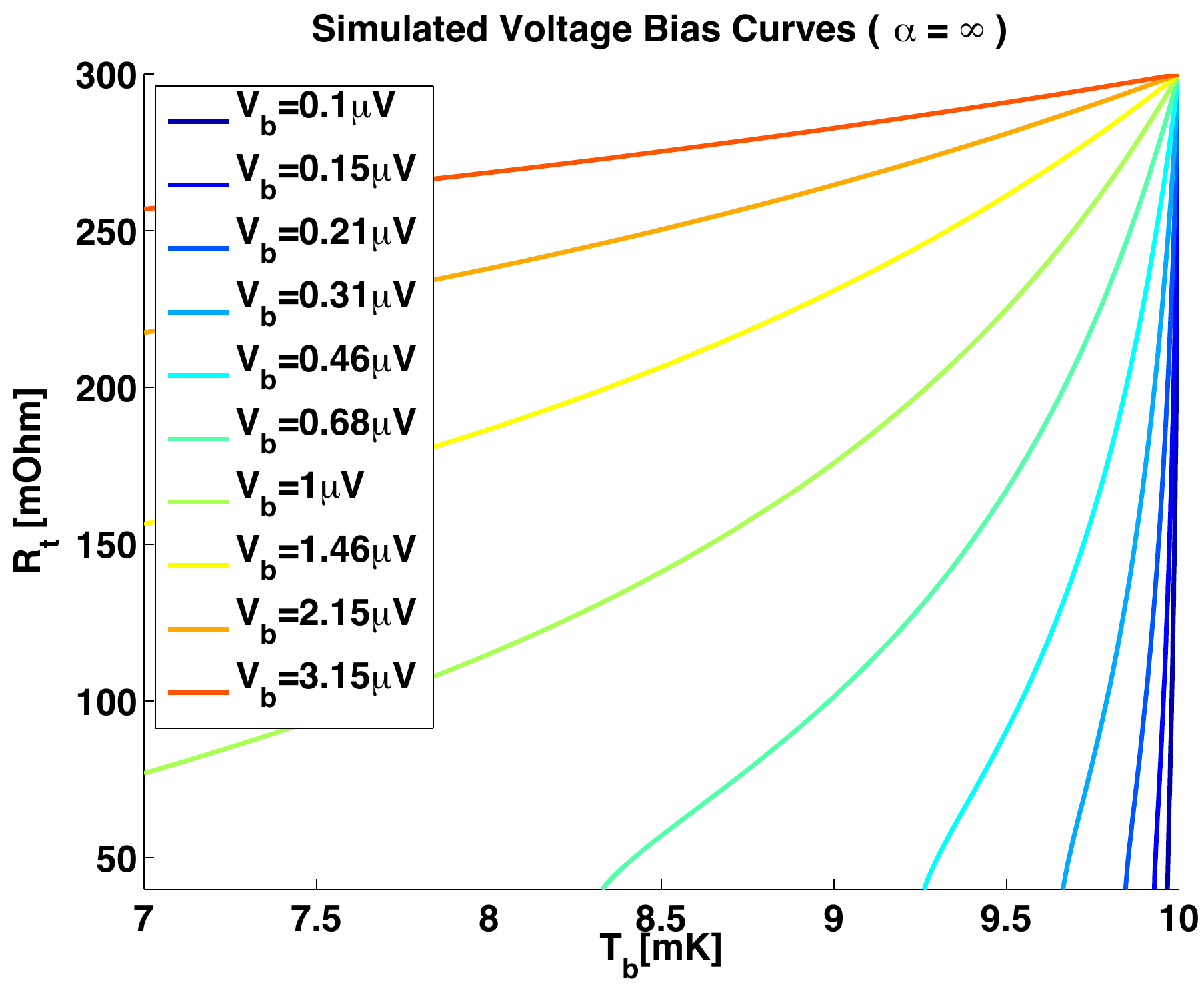}
\caption{\small{Simulated $R_{t}$ vs $T_{b}$ curves for several values of $V_{b}$ for a phase-separated TES with an infinitely sharp transition curve (at $T_{c}=10$~mK) that shows DC sensitivity suppression when biased.}}
\label{fig:RTb_PhaseSep}
\end{figure}

Most importantly, a phase separated TES has significantly suppressed thermal sensitivity when biased. To see this, we note that in the limit of $\alpha =
\frac{T}{\rho}\frac{\partial \rho}{\partial T} \rightarrow \infty$ (i.e. the super conducting transition becoming infinitely sharp and infinitely sensitive to temperature variation), the DC response of the TES can be modeled  analytically by tracking the fractional location of the superconducting/normal transition in the TES, $x$, as a function of $V_{b}$ and $T_{b}$ \cite{circleTES}. For the pertinent case where the electron-phonon coupling along the TES is negligible compared to power flow through the connection to the bath at one end of the TES, the thermal power flowing across the superconducting portion of the TES is constant and can be matched to the joule heating in the normal portion of the TES:
\begin{equation}
{\left(\frac{V_{b}}{R_{l} + R_{n}x}\right)}^{2} R_{n} x= \frac{1}{1-x} \int_{T_{b}}^{T_{c}} \!\! dT \,G_{t\,int}(T)
\label{eq:PhaseSep_equation}
\end{equation}
Using this simplification, we obtain curves of  $R_{t}$ versus $T_{b}$ for a CRESST-like device for several values of  $V_{b}$, as shown in Fig. \ref{fig:RTb_PhaseSep}.

If not for phase separation, all of these curves would be infinitely sharp step functions centered at $T_{b}= T_{c}$. With separation though, we observe significant degradation in device sensitivity that worsens with increasing $V_{b}$.  Basically, a phase separated TES acts as if it has a very large $\beta$ from a DC perspective. This exact behavior is seen in CRESST devices \cite{CRESST_2001_PhononDet,Emilija_2008_thesis}. 
 
Naively, one would think that this suppression in signal amplitude would severely affect phonon power resolution at all frequency scales. However, this is not the case.  At low frequencies, thermal fluctuations across $G_{tb}$  completely dominate the noise in a non-phase-separated TES (Fig. \ref{fig:Spt-Sit} top), and thus both signal and noise are suppressed equally.  Whereas at high frequencies, the total noise is dominated by Johnson noise across the TES and thus the qualitative net effect is a suppression in $\omega_{S/N \, \delta \! P_{t}}$. 

Further, energy that is absorbed by the TES electron system in portions of the TES that are either fully normal or superconducting must first diffuse to the transition region before producing a measurable change in current, which adds an additional strongly expressed diffusive pole into all of the TES transfer functions. In a phase separated CDMS-II TES device for example, measured $\frac{\partial I_{t}}{\partial V_{b}}$ has 2 fall-time poles (plus the $R/L$ pole) with roughly similar weighting  \cite{MCP_Thesis}. Beyond simply adding confusion when trying to understand and model device performance, the most important consequence is to again suppress $\omega_{S/N \, \delta \! P_{t}}$.
 
Finally, due to such large variations in the derivative of the resistivity ($\frac{\partial \rho}{\partial T}$) across the sharp superconducting/normal boundary, thermal power fluctuations within the TES directly couple to the total TES resistance $R_{t}$, and consequently the measured signal current $I_{t}$, giving us an additional noise source with fluctuations on all length scales \cite{circleTES}. Interestingly enough, in discretized simulations of phase separated TES for CDMS-II \cite {MCP_Thesis}, we have found that this additional noise source is largely counter balanced by a suppression in sensitivity to standard TFN noise across $G_{tb}$ at frequencies below the longest diffusive pole, and thus $S_{p\, total}$ primarily sees increases at higher frequencies. 

\section{Bandwidth Mismatch between Signal and Sensor}
\label{sec:BandwidthMismatch}

Before attempting to estimate the sensitivity of CRESST detectors to recoils within the absorber that have both thermal and athermal phonon components, it makes sense to calculate the expected sensitivities of the CRESST detector to the two limiting cases of a Dirac delta thermal energy deposition into the absorber and the TES. These estimates are shown in Table \ref{tab:PhononEres}.  

\begin{table}[h]
\centering
\small
\begin{tabular}{| l | l |}
\hline
$\sigma_{E}$: Eq.~\ref{eq:Eres_lim}				&	1.0 eV \\
\hline
$\sigma_{E_{a}}$: 2D Simulated				&	34 eV (no 1/f)  / 44 eV (1/f)\\
\hline
$\sigma_{E_{t}}$: 2D Simulated					&	0.5 eV (no 1/f)  / 0.5 eV (1/f)\\
\hline
$\sigma_{E_{\gamma}}$ Measured:  ``Julia''  		&      420 eV  \cite{CRESSTII_2009_SplitDetector})\\
\hline
$\sigma_{E_{\gamma}}$ Measured:  Composite  	&      107 eV  \cite{CRESSTII_2014_LowMass})\\
\hline
\end{tabular}
\caption{Simulated and measured CRESST phonon-detector energy resolutions. Note that the simulation does not include the effect of TES phase separation.}
\label{tab:PhononEres}
\end{table}

What is immediately striking is the significantly degraded expected sensitivity for thermal energy deposition in the absorber ($\sigma_{Ea}$) compared to that for direct absorption in the TES ($\sigma_{Et}$). To understand the cause of this suppression factor, we can write $\sigma_{Ea}$ in terms of $\sigma_{Et}$ under the now well motivated assumption that the noise, when referenced to $\delta P_{t}$, should be flat below  $\omega_{S/N \, \delta \! P_{t}}$ (since it is dominated by the naturally flat $G_{tb}$):

\begin{equation}
\begin{split}
\lim_{ \omega_{ta} \ll \omega_{S/N \, \delta \! P_{t}}} \sigma_{Ea}^2 	&= \frac{1}{\int_{0}^{\infty} \frac{d\omega}{2\pi}  \frac{4 |\frac{\partial P_{t}}{\partial E_{a}}(\omega)|^{2}}{S_{P_{t}-total}	(\omega)}}\\
										&\sim \frac{S_{P_{t}\,G_{tb}}(\omega=0)}{\left(\frac{G_{ta,a}}{G_{ta,a} \!+ \!G_{ab}}\right)^{2}\omega_{ta}}\\
										&\sim \sigma_{Et}^2 
														\left(\frac{G_{ta,a} \!+ \!G_{ab}}{G_{ta,a}}\right)^{2}
														\frac{\omega_{S/N \, \delta \! P_{t}}}{\omega_{ta}}										
\end{split}
\label{eq:bandwidth_mismatch}
\end{equation}

Basically, 
 any signal whose bandwidth is smaller than $\omega_{S/N \, \delta \! P_{t}}$ will suboptimally use the sensor bandwidth, leading to poorer resolution as discussed in \cite{Sadoulet_96_BandwidthMismatch}.  To see this in another way, notice that both $\omega_{S/N \, \delta \! P_{t}}$ (Eq.~\ref{eq:w_OF_Pt}) and $S_{p\,total}$ (which is roughly $S_{pt\,G_{tb}}$ as in Eq.\ref{eq:Spt-Gtb}) both scale linearly with $G_{tb}$. Thus, as long as the bandwidth of the signal that is being measured is larger than $\omega_{S/N \, \delta \! P_{t}}$, the baseline energy sensitivity will be independent of the size of $G_{tb}$; increasing $G_{tb}$ increases the noise floor, but this effect is balanced by an increase in sensor bandwidth.  By contrast, when measuring signals with bandwidth below $\omega_{S/N \, \delta \! P_{t}}$ one has accepted a noise penalty but the gained bandwidth is useless.  In summary, because $\omega_{ta} \ll \omega_{S/N \, \delta \! P_{t}}$, the CRESST phonon detector has relatively poor sensitivity to phonon energy that is thermalized within the absorber ($\delta_{Pa}$).

\begin{figure}[h!]
\centering
\includegraphics[width=3.3in]{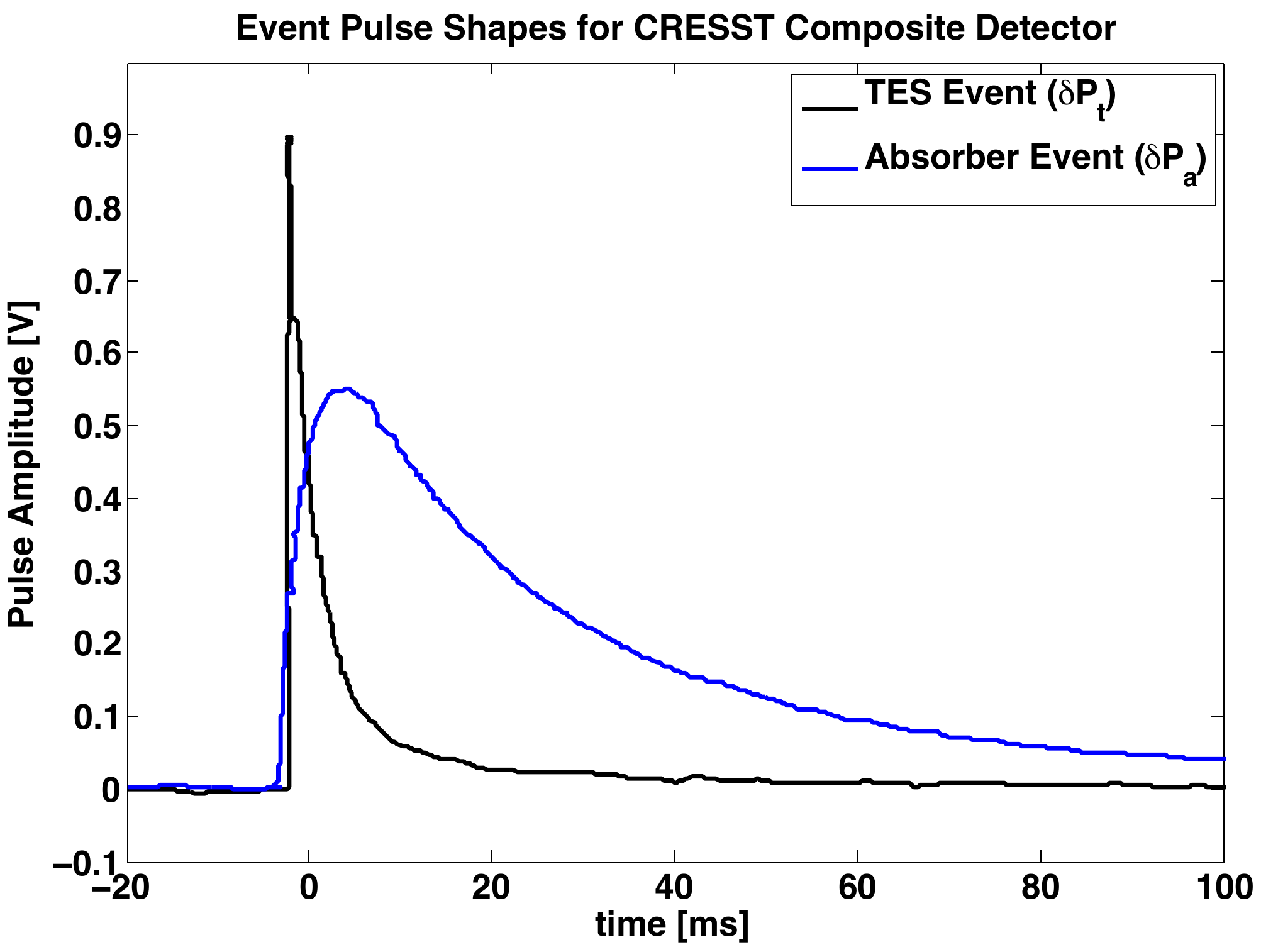}
\caption{\small{Signal pulse shapes for $\gamma$-induced events in the CRESST composite detector ``Rita'' for events that interact in the TES chip (black) and in the absorber attached via an epoxy joint (blue). Clearly, absorber events have significantly suppressed bandwidth and consequently suppressed resolution, as in Eq.~\ref{eq:bandwidth_mismatch}. Data taken from \cite{CRESSTII_2009_SplitDetector}.}}
\label{fig:Traces_CRESST_Composite}
\end{figure}
 
The athermal phonon signal that is thermalized in the TES bypasses the $G_{ta}$ restriction and therefore its bandwidth is not limited by $\omega_{ta}$. However, it still takes time to collect all of the ballistic athermal phonons rattling around in an absorber. In the SuperCDMS athermal phonon iZIP detector for example, this athermal-phonon collection bandwidth $\omega_{nt}$ = 210~Hz while $\omega_{S/N \, \delta \! P_{t}}$= 4~kHz for the 90~mK W TES that CDMS has historically used\cite{MCP_Thesis}.  Thus, they pay a resolution penalty of 5$\times$ due to the bandwidth mismatch between their athermal phonon signal and their sensor bandwidth, completely analogous to the thermalized sensitivity suppression.  

Other examples of sensitivity suppression due to poor signal /sensor bandwidth matching can be found in the new CRESST composite detectors (Fig. \ref{fig:Traces_CRESST_Composite}) \cite{CRESSTII_2009_SplitDetector} as well as the AMORE MMC based calorimeters \cite{AMORE_2014}. In the CRESST composite detector, the athermal phonon signal from an event in the absorber crystal must first be transmitted across an epoxy glue joint between the absorber crystal and the substrate on which the TES is fabricated, yielding $\omega_{nt} \sim$7 hz  (fall time $\sim$ 23 ms). If we associate the fall-time pole of the phonon signals for events hitting the TES chip substrate ($\sim$ 50 Hz) with $\omega_\mathit{eff}$ and we further recognize that $\omega_{S/N \, \delta \! P_{t}} > \omega_\mathit{eff}$ for CRESST detectors (since they purposely bias to minimize electro-thermal feedback), then we can deduce that the new CRESST composite has a signal/sensor bandwidth suppression factor that is $>$10$\times$.  In fact, using our 2D simulation, we estimate that this bandwidth-limited energy impulse into the TES, $\sigma_{E_{t}}(\omega_{nt} \! =\! \mathrm{50\,Hz})$, would have a resolution of 5 eV, 10$\times$ worse than the Dirac-Delta sensitivity.
 
 In summary, our first and most important very low-temperature detector design objective is that the signal bandwidth must be larger than the sensor bandwidth, a design goal that, to our knowledge, has not been accomplished by any very low-temperature large-mass detector but that is standardly implemented in 100 mK TES calorimeters for x-ray \cite{Saab_06_xray} and $\gamma$ \cite{Irwin_GammaTES} applications. Please note that there is some subtlety in this design rule. CRESST detectors use pulse-shape differences between absorber and TES events to suppress TES chip backgrounds (Fig. \ref{fig:Traces_CRESST_Composite}); they have both an energy and a discrimination signal with different bandwidths. Thus, the optimal strategy is likely to choose $\omega_{ta} \sim 1/2 \omega_{S/N \, \delta \! P_{t}}$ so that the device discrimination threshold is equal to its energy threshold.

Secondly, if we are able to design very low-temperature calorimeters that use the entire unsuppressed sensor bandwidth (design rule 1), we will become directly sensitive to TES phase separation. Consequently we impose a second design constraint that $G_{tb} \ll G_{t \, int}$. As an added benefit, we think device operation and analysis will be significantly less complex, because we also naively expect that phase-separated TESs are highly sensitivity to position-dependent variations in $T_{c}$ and other thin-film properties that are likely culprits for at least some of the non-linearities that have been found when biasing CRESST devices \cite{Emilija_2008_thesis}.

\section{Expected CRESST Energy Resolution \& Athermal Phonon Collection Efficiency }
\label{sec:PhononCollection}

We have shown that the infinite bandwidth $\sigma_{Et}$ in Table \ref{tab:PhononEres} significantly overestimates the athermal phonon energy sensitivity, since it does not account for the athermal phonon collection bandwidth, $\omega_{nt}$. Additionally, we must account for the fact that only a fraction $\epsilon_{nt}$ of the athermal phonons are collected in the TES, and thus a particle recoil in the absorber should be modeled as
\begin{equation}
\begin{split}
	\frac{dP_{t}}{dE_{\gamma}} &= \frac{\epsilon_{nt}}{1+j\omega/\omega_{nt}}  \\
	\frac{dP_{a}}{dE_{\gamma}} &= 1-\epsilon_{nt}	
\end{split}
\label{eq:dIdExyz}
\end{equation}
For an upper bound on $\epsilon_{nt}$, we note that optical photons directly incident on a W TES have been measured to deposit $\sim$ 42\% of their total phonon energy into the W electronic system \cite{Burney_2006_OpticalTES}. Another guide to the size of $\epsilon_{nt}$ comes from the SuperCDMS iZIP athermal phonon detector, which measures a total athermal phonon collection that is 10--20\% of the total deposited energy~\cite{MCP_Thesis}. Using a reasonable (but certainly not measured) value of $\epsilon_{nt}=20\%$, we estimate a $\sigma_{E_{\gamma}}$ = 15 eV.  Although this is 7$\times$ better than the best achieved resolution in a CRESST composite detector (107 eV), it is also 30$\times$ worse than the naive expectation of 0.5 eV, and thus we believe that sensor/signal bandwidth mismatch and non-unity athermal phonon collection efficiency are the dominant sensitivity degradation mechanisms. 
\section{Parasitic Power Loading}
\label{sec:ParPower}

\begin{figure}[htbp!]
\centering
\includegraphics[width=2.9in]{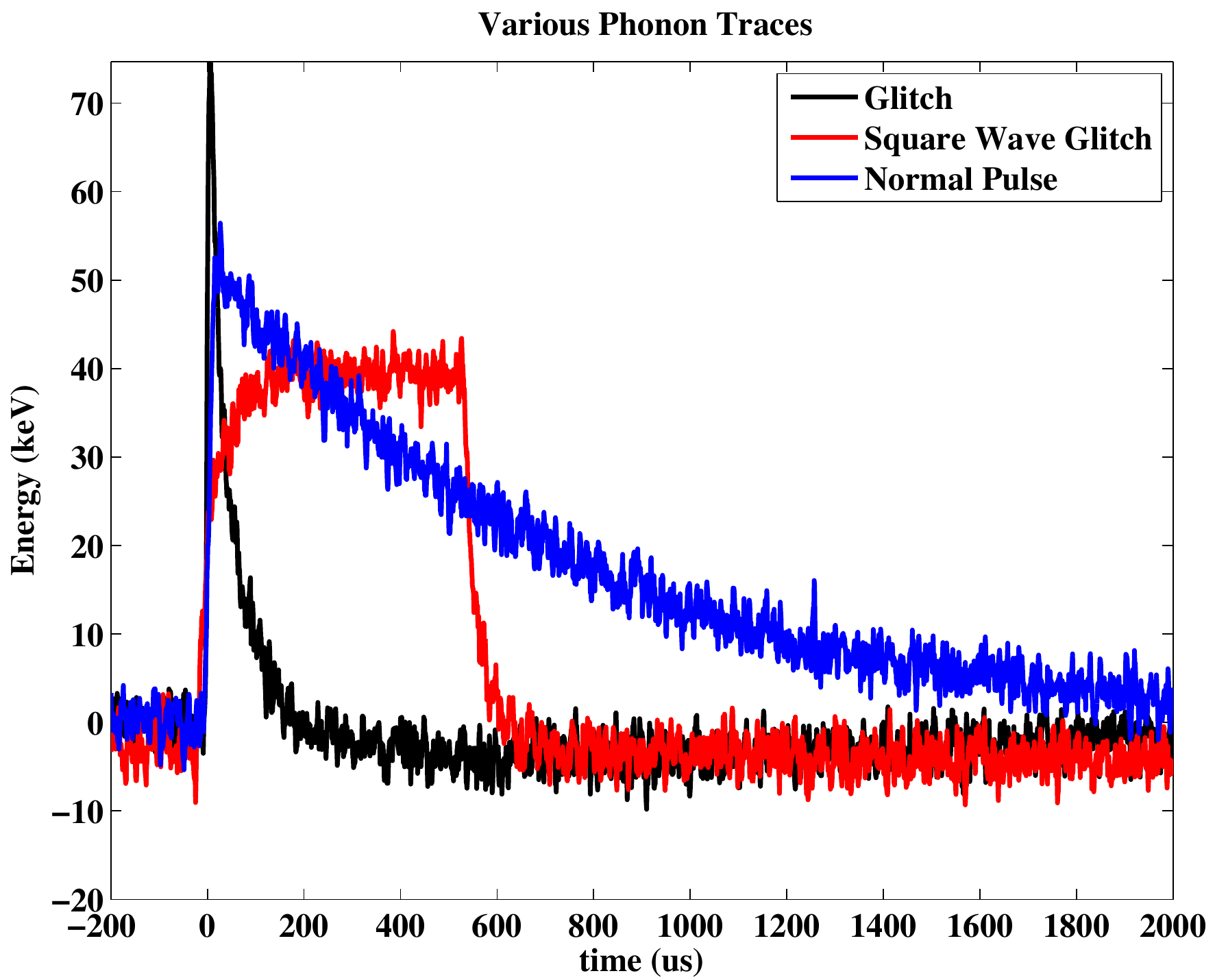}
\caption{\small{Environmental noise pulses (red and black) illustrate that SuperCDMS detectors are strongly susceptible to direct TES heat by parasitic power noise.}}
\label{fig:GlichPulse}
\end{figure}

Both design drivers so far discussed suggest that to become more sensitive an optimized very low-temperature calorimeter should have lower $G_{tb}$ than found in the CRESST phonon-detector design. One negative consequence of this change is greater sensitivity to parasitic heating. This is clearly a concern for SuperCDMS because their devices produce TES heating signals due to cell phone usage in the laboratory (fortunately with a very distinct pulse shape from actual events, as shown in Fig. \ref{fig:GlichPulse}), and thus the old CDMS-II electronics do not adequately shield their TESs from high-frequency environmental noise. Although this is currently only an interesting nuisance, it highlights the parasitic power problem that SuperCDMS faces.  As they lower G$_{tb}$ in their own devices, this and other parasitic power sources heat the detector to a greater and greater degree, eventually driving the TES normal and rendering it inoperable. Lower $G_{tb}$ devices will require better environmental noise shielding.

\begin{figure}[htbp!]
\centering
\includegraphics[width=2.9in]{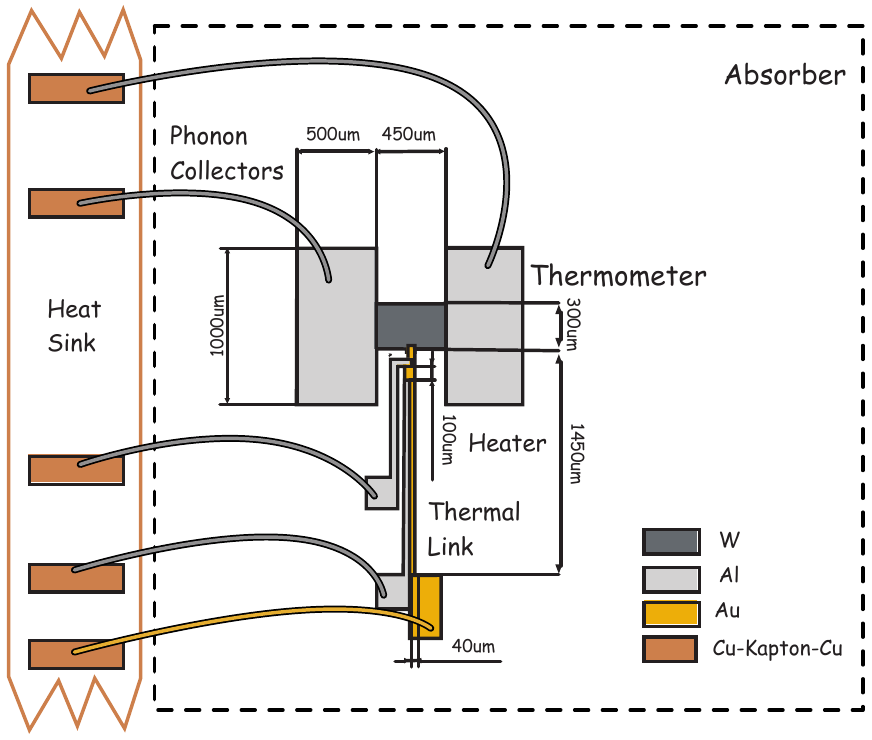}
\caption{\small{TES Sensor Geometry for CRESST-II light detector \cite{Emilija_2008_thesis}}}
\label{fig:LightDet_diag}
\end{figure}

Thus, it is reasonable to question if DC parasitic power noise necessitates the use of large $G_{tb}$ in CRESST phonon detectors, thereby breaking our first 2 design rules and lowering detector sensitivity. Further, if so, are all very low-temperature calorimeters similarly limited? We do not believe this to be this case for 2 reasons. First, in the very same cryostat and with identical electronics, CRESST also operates light detectors that are Si or SOS wafers instrumented with the TES design shown in Fig. \ref{fig:LightDet_diag}.  Notice that for these detectors, the TES is not directly connected to the bath through an Au wirebond. Instead the coupling is through a fabricated Au thin film impedance with an estimated $G_{tb}$= 21.1 pW/K for a $T_{c}$ = 10mK which is 350$\times$ smaller than the $G_{tb}$ of the Au wirebond used on their phonon detectors. Consequently, if DC parasitic power noise was even remotely problematic for the CRESST phonon detector, their light detector would be inoperable.

Of course, when trying to assess the validity of lowering $G_{tb}$ during the optimization of TES-based light detectors (something we also discuss due to applicability for both dark matter and double-beta decay experiments), the above explicitly pertinent observation is insufficient.  Fortunately, however, there has recently been enormous emphasis on designing TES-based sensors for space-born infrared spectrometry by SPICA/SAFARI, and they have been able to achieve parasitic power loads of $\sim$2 fW, 25$\times$ less than the estimated bias power, $P_{to}$, for the current CRESST light detector and 10$^{4}\times$ less than their phonon detector \cite{Goldie_2011_SPICATES}. Consequently, achieving parasitic power loads that are much less than $P_{to}$ for low G$_{tb}$ devices may be difficult but should be possible.

\section{Limits on Sensor Bandwidth due to Event Rate}
\label{sec:EventRate}
Unlike in 100 mK x-ray, $\gamma$ and $\alpha$ calorimeters, the extremely low background and signal rates expected in underground rare-event and exotic-decay searches have meant that pileup (pulses from distinct interactions overlapping due to high rates and finite bandwidth) rejection and the ability to handle large event rates have not been primary design drivers.

In the future, however, this requirement will become much more constraining for some applications. For example, the most daunting pileup requirements come from reactor-sourced coherent neutrino scattering experiments because the cosmic background is quite large due to minimal rock overburden. As a first very rough estimate of the necessary start-time sensitivity, we note that the original CDMS shallow site experiment had a similar overburden and used an anti-coincidence window of 25 $\mu$s for activity in their plastic-scintillator muon veto \cite{CDMS_SUF_02}.  Since the optimally estimated start-time resolution (assuming no pulse-shape dependence) is 
\begin{equation}
\sigma_{t_{o}}^2=\frac{1}{E^{2}} \frac{1}{\int_{0}^{\infty} \frac{d\omega}{2\pi}  \frac{4 \omega^{2} |\frac{dI_{t}}{dE_{\gamma}}(\omega)|^{2}}{S_{I \, tot}(\omega)}}
\label{eq:to_res}
\end{equation}
 which can be simplified to
\begin{equation}
\sigma_{t_{o}} \sim  \frac{1}{\omega_{S/N \, \delta \! P_{t}}} \frac{\sigma_{E}}{E}
\label{eq:to_res2}
\end{equation}
when flat noise is assumed and $\omega_{S/N \, \delta \! P_{t}}$ is the lowest pole of the design, we estimate that a reactor-sourced coherent neutrino scattering experiment needs an $\omega_{S/N \, \delta \! P_{t}}$ of 1.3 khz to have the requisite start-time resolution for near-threshold events ($E = 10 \sigma_{E}$). Unfortunately, such a large $\omega_{S/N \, \delta \! P_{t}}$ is simply inconsistent with our other design requirements for thermal calorimeters. Consequently, we believe that this application is better suited to an athermal-only phonon detector design, as typified by a SuperCDMS detector. 

Pileup rejection has also become one of the dominant design drivers for LUCIFER and other high-Q$_{\beta \beta}$ neutrinoless double-beta decay experiments because they expect the background in their signal region to be dominated by un-vetoed pileup of lower energy 2 $\nu$ double-beta decay events for which pileup rejection is assumed to be ineffectual below 5 ms \cite{CUORE_ZnMoO4}. Luckily, with such a large signal, start-time resolutions of this order are achievable (see Sec. \ref{sec:DBDdet}) and thus, if required, this is a reasonable design requirement.

\section{Ease of Fabrication \& Cosmogenic Background Suppression: Separated TES Chip}
\label{sec:DeviceGeom}

CUORE and EDELWEISS have continued to use NTD sensor technology,  despite the many benefits offered by TES readout, including:
\begin{enumerate}
	\item A TES is fundamentally more sensitive than a NTD (larger $\alpha$);
	\item The SQUIDs used in TES readout have lower $1/f$ noise than JFETs or HEMTs used for first stage amplification for NTDs;
	\item  Low-impedance sensors (TESs) are fundamentally less sensitive to vibrationally induced capacitance changes in readout compared to high-impedance sensors (NTDs); and
	\item SQUIDs have significantly lower heat loads than the JFETs or HEMTs used in NTD readout, and can therefore be placed significantly closer to the detector, simplifying electronics and cryostat design.
\end{enumerate}

One reason for this is that both TES-based massive-detector groups (CRESST and CDMS) fabricated their TESs directly upon the absorber surface, a feat that required enormous fabrication process R\&D since every facet of standard microprocessor fabrication (photolithography, etching, thin film deposition) had to be retrofitted for thick and massive substrates. In fact, even with over a decade of R\&D, fabrication yields were still a significant resource drain until recently on CDMS. Further, the time and labor intensive nature of microprocessing fabrication means that absorbers spend a significant amount of time on the surface being cosmogenically activated, certainly a disadvantage for low mass WIMP searches, for example. This direct absorber fabrication is required since the W TES in CRESST has 2 distinct functions.  First, it is a temperature sensor.  This is the functionality that requires microprocessor fabrication techniques on multiple different thin film layers: Al for the superconducting bias rails, W for the TES, and Au for the thermal connection to bath. Second, the electron-phonon coupling within the W film acts as $G_{ta}$. Because it is only this latter functionality that requires fabrication directly upon the absorber, design goals of fabrication simplicity and minimum cosmogenic exposure require that the TES does not act as the thermal link between the sensor and the absorber. 

\begin{figure}[h!]
\centering
\includegraphics[width=3.3in]{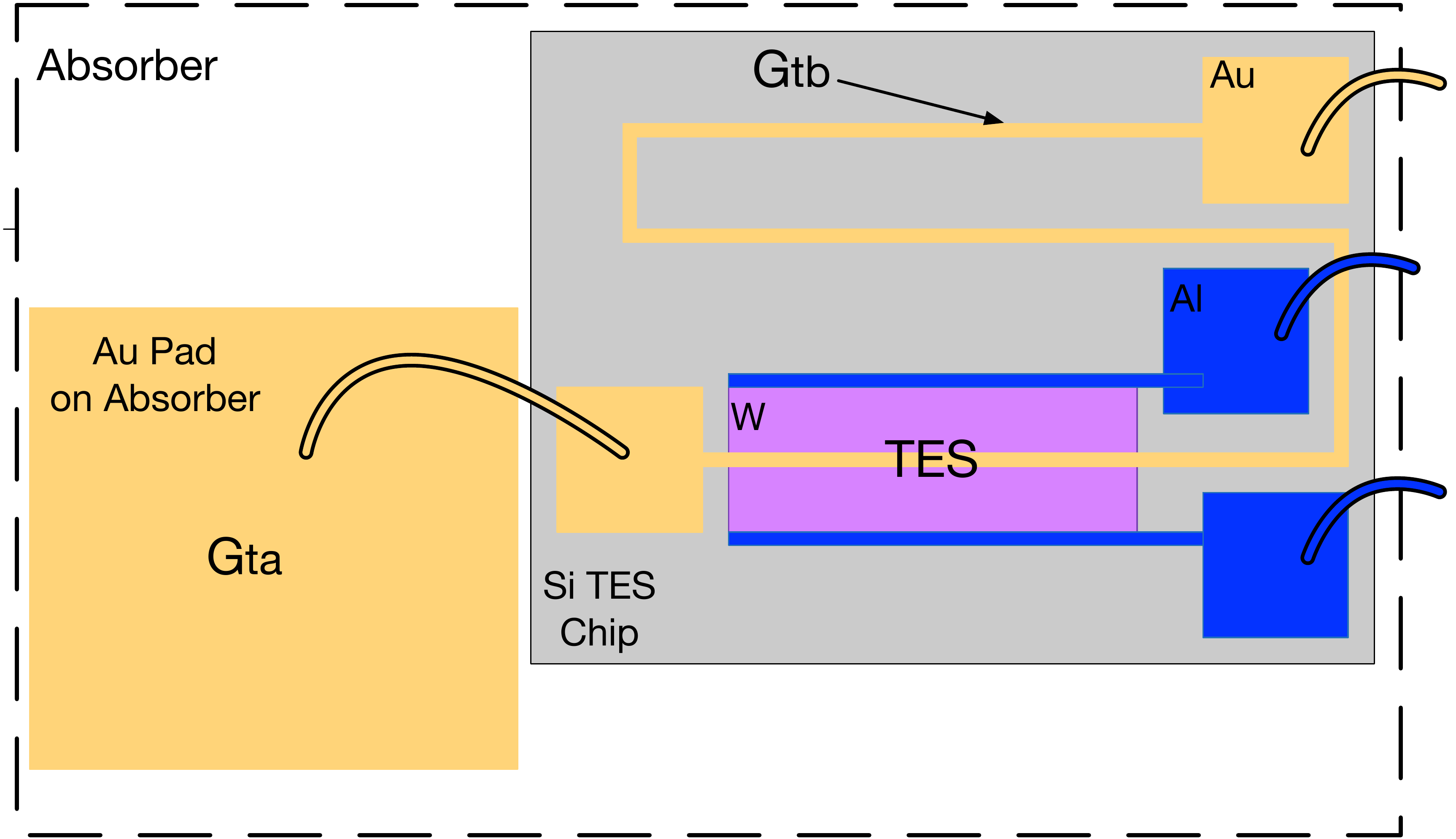}
\caption{\small{Optimized large-mass calorimeter sensor design. Only the large Au pad is directly fabricated on the large absorber.}}
\label{fig:TESchip}
\end{figure}

As illustrated in Fig. \ref{fig:TESchip}, we propose to deposit a large, single layer Au thin-film pad directly onto the large absorber substrates that plays the role of $G_{ta}$. This can be done using only shadow mask techniques (albeit with a depostion machine modified for thick substrates) and consequently fabrication should have very high yield and be relatively hassle free since there is no photolithography and etching. As an added benefit (in fact, perhaps the most important benefit),
this permits use of any metal rather than being constrained to W; we choose Au which has an order of magnitude larger electron-phonon coupling than W for a given thermal capacitance.

The fabrication intensive TES can then be separately fabricated on standard thin substrates for fabrication ease, where the material chosen is not necessarily identical to that of the large absorber (Si, Ge, Al$_{2}$O$_{3}$, CaWO$_{4}$). Further, each and every 100 mm wafer can produce over 20 devices. Thus, device fabrication throughput could easily be 80$\times$ that of CDMS (in the standard SuperCDMS fabrication procedure, 4 fully processed test wafers are produced for every detector). This physical separation also allows for testing of the TES sensor die above ground before connection to the absorber.  This has significant advantages:  the cost savings of sensor testing above ground rather than in an underground laboratory is substantial; and one can choose only the best sensors to match with expensive absorbers, particularly useful in the case of double-beta decay enriched crystals.

Thermal connection between the TES and the G$_{ta}$ absorber pad is then accomplished via Au wire bonding, while the simpler mechanical connection can be accomplished with epoxy in an arbitrary (and somewhat haphazard) manner; without any thermal-conductance requirements, very small-area single dot epoxy joints are possible that do not mechanically stress the absorber \cite{CRESSTII_2009_SplitDetector} ). Another possibility is that the TES chips are mechanically supported by the detector housing. Of course, the heat capacity of the Au wirebond between the TES and the absorber ($\sim$1 pJ/K at 10 mK) is entirely parasitic and is simply the price paid for the ease of fabrication. Most importantly, the thermal conductance of both the internal pad, $G_{pad\,int}$, and that of the Au wirebond, $G_{bond\,int}$, must be much larger than $G_{tb}$ so as to satisfy the bandwidth design rules discussed in Sec. \ref{sec:BandwidthMismatch}.

It is interesting to reiterate how similar and yet how distinct this design is from the CRESST composite detector \cite{CRESSTII_2009_SplitDetector}. On the one hand, this design is clearly derivative. As with the CRESST composite detector, the TES is fabricated on a separate and much thinner substrate that drastically simplifies fabrication (and in the CRESST case, improves scintillation yield in the primary absorber).  On the other hand, the difference is profound. In the CRESST design, energy transport between the 2 systems is accomplished via phonon transport through an epoxy joint that significantly suppresses both athermal and thermal signal bandwidths (making our bandwidth design objectives very difficult to achieve). Further,  low-impedance phonon transport via epoxy couplings requires large cross-sectional areas that make the detector much more prone to crack via differential thermal contraction among the epoxy, the absorber substrate, and the TES chip substrate. The CRESST composite detector mitigates this problem somewhat by using CaWO$_{4}$  for both the TES substrate and the absorber.  However, this limits possible detector materials.  NaI, for example, is an attractive material to search for spin or orbital angular momentum coupling dark matter \cite{Anand_2014_EFTDM}. Furthermore, scintillating NaI calorimeters would directly test the anomalous annual modulation signal observed in DAMA \cite{Nadeau_2014_NaI}. Unfortunately though, NaI is very hygroscopic and thus any complex photolithography is impossible.

In summary, this design should retain all the benefits of TES performance and yet also have the fabrication simplicity, cosmogenic benefits, and material freedom of  NTD based detectors.

This device design should also be critically compared to the AMORE MMC calorimeter design \cite{AMORE_2014}. Removing the trivial difference of the use of an MMC instead of a TES sensor element, all pertinent design rules that we have developed were followed with the single and very important exception that their sensor bandwidth is much larger than their signal bandwidth ($\omega_{ta}$). This choice suppresses their zero energy sensitivity but even more importantly drastically increases their sensitivity to position dependence (Secs. \ref{sec:PosDep} and \ref{sec:DBDdet}), a fact that severely limits the viability of their current devices. 

\section{Position Dependence Requirements in Neutrinoless Double Beta Decay Searches}
\label{sec:PosDep}
Because of the approximately exponential shape of the WIMP nuclear-recoil energy spectrum, a slight position or temporal systematic of $<$5\% on the recoil energy estimate in a CRESST phonon detector does not significantly affect their sensitivity. 
By contrast, a neutrinoless double-beta decay experiment like CUORE/LUCIFER \cite{CUORE_ZnMoO4} requires the maximum achievable energy resolution \emph{at 3 MeV}, and thus any variation in the pulse shape or magnitude of the signal due to event location in the absorber must be minimized in addition to having excellent baseline energy sensitivity. Specifically, a well designed double beta detector should be limited by systematics in the absorber thermalization process, with $\frac{\Delta E}{E}$ measured to be 5$\times$10$^{-4}$ in TeO$_{2}$ \cite{CUORE_TeO2_QF}) and 2x10$^{-3}$ in ZnMoO$_{4}$ \cite{CUORE_ZnMoO4}.


\begin{figure}[h!]
\centering
\includegraphics[width=3.3in]{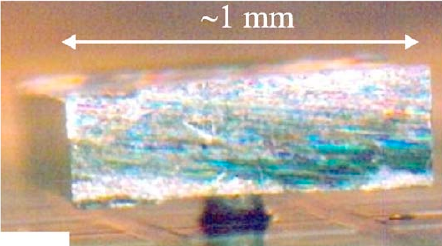}
\caption{\small{ NIST $\gamma$ calorimeter with $\sigma$= 21 eV at 103 keV using a TES with T$_{c}\sim$ 100 mK. \cite{Irwin_GammaTES}}}
\label{fig:GammaDet}
\end{figure}

To guide us in developing design rules to minimize position dependence, let us look more carefully at the 100\,mK TES $\gamma$ calorimeter shown in Fig. \ref{fig:GammaDet} that achieved a $\frac{\Delta E}{E}<$5$\times$10$^{-4}$\cite{Irwin_GammaTES}. First and foremost, the only thermal and structural link between the absorber and the bath goes through the TES ($G_{ab}$=0). To illustrate the importance of this for suppressing position dependence, consider the design in \cite{Tali_11_CNS} where the primary thermal path from the TES to the bath is through the absorber ($G_{tb}=0$, $G_{ab}\neq0$). In this case, athermal phonons produced in the initial interaction will preferentially thermalize in either the TES ($G_{tb}$) or in the metal pad thermally linked to the bath ($G_{ab}$), depending on the event's location in the absorber and leading to significant unwanted position dependence in $\delta I_{t}$. 

Unfortunately, $G_{ab}$ free designs are significantly more difficult to achieve when the absorber weighs $\mathcal{O}(kg)$ rather than $\mathcal{O}(g)$, and thus this will be an aspirational goal rather than a requirement. We will solely require that $G_{ab}\ll G_{ta}$, and that any $G_{ab}$ be diffusive rather than ballistic so that non-thermalized excitations would find it difficult to escape through $G_{ab}$. 



Secondly, the size of the diffusive-phonon thermal link between the absorber and the TES in this $\gamma$ calorimeter was chosen to be much slower than the thermalization/position-dependent time scale, $\omega_{nt} \gg \omega_{ta} \gg \omega_{S/N \, \delta \! P_{t}}$.  This choice means that unwanted position information with frequencies around $\omega_{nt}$ have 2 pole suppression over many decades in their current response. 

Unfortunately, thermalization is very slow in the insulator and semiconductor absorbers used in massive calorimeters compared to the superconductors and metal absorbers used in $\gamma$ and x-ray calorimeters. SuperCDMS iZIP detectors, for example, see variations in athermal phonon power absorbed by different channels on the same crystal up to 300 $\mu$s after an event interaction. Such long position-dependent thermalization times put downward pressure on both $\omega_{ta}$ and $\omega_{S/N \, \delta \! P_{t}}$ when using this bandwidth design scheme. CUORE actually attempts to do exactly this in their NTD based calorimeter by using a small dot of epoxy as $G_{ta}$, with the hope of suppressing any non-thermalized position-dependent phonon signal from reaching their TES.  Unfortunately, this choice in combination with poor electron-phonon coupling within the NTD means that $G_{ab} > G_{ta}$, and thus most of the absorbed energy is not measured by the NTD but directly shunted to the bath, suppressing baseline energy sensitivity and increasing DC susceptibility to position dependence \cite{CUORE_ThermalConductance_92}.

In our proposed calorimeter design, the above bandwidth design scheme is very difficult to achieve because the dominant phonon thermalization mechanism is through electron-phonon coupling within the metal pad that acts as $G_{ta}$; $\frac{\partial P_{t}}{\partial E_{\gamma}}$ (Eq.~\ref{eq:dItdE}) will always be non-zero. Thus, we must be content with single-pole suppression of position dependence. On the otherhand, since we now control the rate of athermal phonon thermalization ($\omega_{nt}$), we can maximize this single-pole suppression by making the pad as large as possible (i.e. set the design requirement that $\omega_{nt} \,  \& \, \omega_{ta} \gg \omega_{S/N \, \delta \! P_{t}}$). As an aside, note that suppression of position dependence via pole suppression is incompatible with the pulse-shape discrimination between TES chip events and absorber events that CRESST enjoys because it requires ($\omega_{ta} \sim$ \sfrac{1}{2} $\omega_{S/N \, \delta \! P_{t}}$).

As a rough but conservative estimate of the position-dependent systematics in both the energy and start-time estimators, we assume that the thermal power signals into the TES and absorber can be modeled as
\begin{equation}
\begin{split}
	\frac{dP_{t}}{dE_{\gamma}} &= \epsilon \\
	\frac{dP_{a}}{dE_{\gamma}} &= 1-\epsilon	
\end{split}
\label{eq:dIdExyz}
\end{equation}
where we vary $\epsilon$ from 0--10\%, using SuperCDMS position dependence as a rough guide (we have pushed $\omega_{nt} \rightarrow \infty$ to maximally accentuate the position dependencies). 

\section{Design Sketch of 1.75kg $\text{ZnMoO$_{4}$}$ Detector for Neutrinoless Double Beta Decay} 
\label{sec:DBDdet}

In Table. \ref{tab:OptCUORE} we flesh out the specifications and simulated performance for a 1.75 kg ZnMoO$_{4}$ device that follows all of the design rules so far discussed. Such a massive absorber (or the addition of parasitic heat capacitance as was done in \cite{alphaTES_08}) is required to keep $Q_{\beta \beta}$ within the dynamic range of the TES. We recognize that the use of such large crystals increases the need for pileup rejection (i.e. timing requirements) \cite{CUORE_ZnMoO4}. 

\begin{table}
\centering
\small
\begin{tabular}{| l | m{2.8cm} | m{2.4cm} | l |}
\hline
\multicolumn{4}{| c |}{ZnMoO$_{4}$ absorber} \\
\hline
$V_{a}$	& Volume								&														& $\pi$(40mm)$^{2}$x80mm \\  
$T_{ao}$	& Operational temperature					&														& 9.33 mK \\ 
$C_{a}$	& Heat capacity 						& $\Gamma_{\text{\tiny{ZnMoO$_{4}$}}} V_{a} T_{ao}^{3}$				& 191 $\mathrm{\frac{pJ}{K}}$\\ 
\hline
\multicolumn{4}{| c |}{Au thin film thermal link between  TES and bath} \\
\hline
$l_{tb}$	& Length								&														& 900 $\mu$m  \\ 
$A_{tb}$	& Cross sectional area					&														& 10 $\mu$m x 300 nm \\ 
$G_{tb}$	& Thermal conductance from TES to bath 	& $\frac{v_{f} d_{e}}{3} \Gamma_{Au} T_{to} \frac{A_{tb}}{l_{tb}}$  & 286 $\mathrm{\frac{pW}{K}}$\\ 
\hline
\multicolumn{4}{| c |}{Au thermalization film on absorber (p1)} \\
\hline
$V_{p1}$	&  Volume								& 														& 4$\pi$(9mm)$^{2}$x300nm\\ 
$C_{t\,p1}$& p1 component of TES heat capacity 		& $\Gamma_{Au} V_{p1} T_{to}$								&187 $\mathrm{\frac{pJ}{K}}$ \\ 
$G_{ta}$	& Thermal conductance between TES and absorber %
											& $n_{ep} \Sigma_{ep} V_{p1}T_{to}^{n_{ep}-1}$				        &37 $\mathrm{\frac{nW}{K}}$\\ 
\hline
\multicolumn{4}{| c |}{Au wirebond between  TES and absorber} \\
\hline
$l_{w}$		& Length							&														& 1.0 cm \\ 
$A_{w}$		& Cross sectional area				&														& 4$\pi$(7.5$\mu$m)$^{2}$\\ 
$G_{w}$		& Thermal conductance 				& $\frac{v_{f} d_{e}}{3} \gamma_{Au} T_{to} \frac{A_{w}}{l_{w}}$	& 21.6 $\mathrm{\frac{nW}{K}}$\\ 
$C_{t \, w}$	& Bond component of TES heat capacity &														& 4.6 $\mathrm{\frac{pJ}{K}}$\\ 
\hline
\multicolumn{4}{| c |}{W  TES} \\
\hline
$A_{t}$			& Cross sectional area					&														& 3.5 mmx150 nm \\  
$l_{t}$			& Length								&														& 700 $\mu$m \\ 
$T_{c}$			& Transition temperature					&														& 10 mK \\ 
$\Delta T_{90-10}$	& Transition width 						& (T$_{90\%}$-T$_{10\%}$)									& 2 mK \\ 
$T_{to}$			& Operating temperature					&														& 9.33 mK \\  
$I_{to}$			& Operating current						& 														& 5.4 $\mu$A \\ 
$C_{t \, W}$ 		& W component of TES heat capacity		& $f_{sc} \Gamma_{W} V_{t} T_{to}$								& 0.96 $\mathrm{\frac{pJ}{K}}$ \\ 
$G_{t\,int}$		& Internal TES conductance				& $\frac{L_{wf}}{\rho_{W}} \frac{A_{t}}{l_{t}/2} T_{to}$					& 4.9 $\mathrm{\frac{nW}{K}}$\\  
$R_{n}$			& Normal resistance					& $\rho_{W} \frac{l_{t}}{A_{t}}$									& 100 $m\Omega$\\ 
$R_{to}$ 			& Operating point						& R$_{n}$/5												& 20 $m\Omega$\\ 
$\alpha$			& Thermal sensitivity					& $\frac{T_{to}}{R_{to}} \left. \frac{\partial R}{\partial T_{t}} \right|_{I_{t o}}$ & 16.4 \\ 
$\beta$ 			& Current sensitivity						& $\frac{I_{to}}{R_{to}} \left. \frac{\partial R}{\partial I_{t}} \right|_{T_{t o}}$ & 0.05\\ 
\hline
\multicolumn{4}{| c |}{ Dynamical time constants} \\
\hline
$\tau_{L/R}$		& Squid inductor time constant			&																& 15 $\mu$s \\ 
$\tau_\mathit{eff}$		& sensor fall time					&																& 365 ms\\ 
$\tau_{ta}$		& Absorber/TES coupling time constant 	&																& 2.6 ms\\	
\hline
\multicolumn{4}{| c |}{ Estimated resolution and saturation energy} \\
\hline
$\sigma_{E}$		& Estimated baseline energy resolution	 &																& 3.54 eV\\ 
$\sigma_{to}$ 		& Estimated timing  resolution @ 3 MeV  &																& 170 $\mu$s\\ 
$\Delta E / E$		&Position dependence 				 &																& 6x10$^{-4}$ \\ 
$E_{sat}$			&Absorber saturation energy 			 &																& 4.1 MeV \\ 
\hline
\end{tabular}
\caption{Optimized 1.75kg ZnMoO$_{4}$ detector for next-generation double-beta decay experiments}
\label{tab:OptCUORE}
\end{table}

The Au pad volume on the absorber was chosen to be quite large so that $\omega_{ta} \gg \omega_{S/N \, \delta \! P_{t}}$. Notice that even with such a relatively large volume, the parasitic heat capacity of the pad is only $\sim$ 80\% that of the ZnMoO$_{4}$. The thickness of the pad at 300 nm was a compromise. On the one hand we wanted to cover the the largest possible surface area, so as to maximize $\omega_{nt}$.  On the otherhand, we want the internal thermal conduction of the pad $G_{pad\, int} \gg G_{ta}$. Since the parasitic heat capacity of an individual Au wire bond is also relatively small (1.2~pJ/K) compared to the crystal, we used 4 and assumed an extra long 200 $\mu$m bond tail at both ends for greater thermal conductance. 

For the TES itself, we chose a W TES with an $R_{n}$=100~m$\Omega$ that we would run low in the transition (20\% R$_{n}$) to maximize the dynamic range. Further, we would use R$_{l}$= 3~m$\Omega$ to be in the strong electro-thermal limit. This aspect ratio was chosen so that $G_{t\,int} \gg G_{tb}$ and could potentially be further lowered. Since the heat capacity of the TES is subdominant compared to both the crystal and the $G_{ta}$ pad, the sole constraints on its size are related to sensor performance related. Since $\beta$ increases with current density (which scales as V$_{t}^{-1/2}$, where V$_{t}$ is the TES volume), we will choose the overall TES size such that the current density is 1/2 that found in the current CRESST detectors. The use of low-resistivity Ir/Au or Mo/Au bilayers \cite{CRESST_1995_ProbstModel,Damayanthi_06_IrAuTES,SmithMoAuTES} is also certainly possible and would suppress TES inhomogeneities but with the potential for larger $\beta$. 

Finally, $G_{tb}$, is a thin Au film impedance similar to that used by CRESST in their original light detectors \cite{Emilija_2008_thesis}. Its size is set so that $\omega_{S/N \, \delta \! P_{t}}$ is as small as allowed by 1/f noise constraints.

\begin{figure}[h!]
\centering
\includegraphics[width=2.9in]{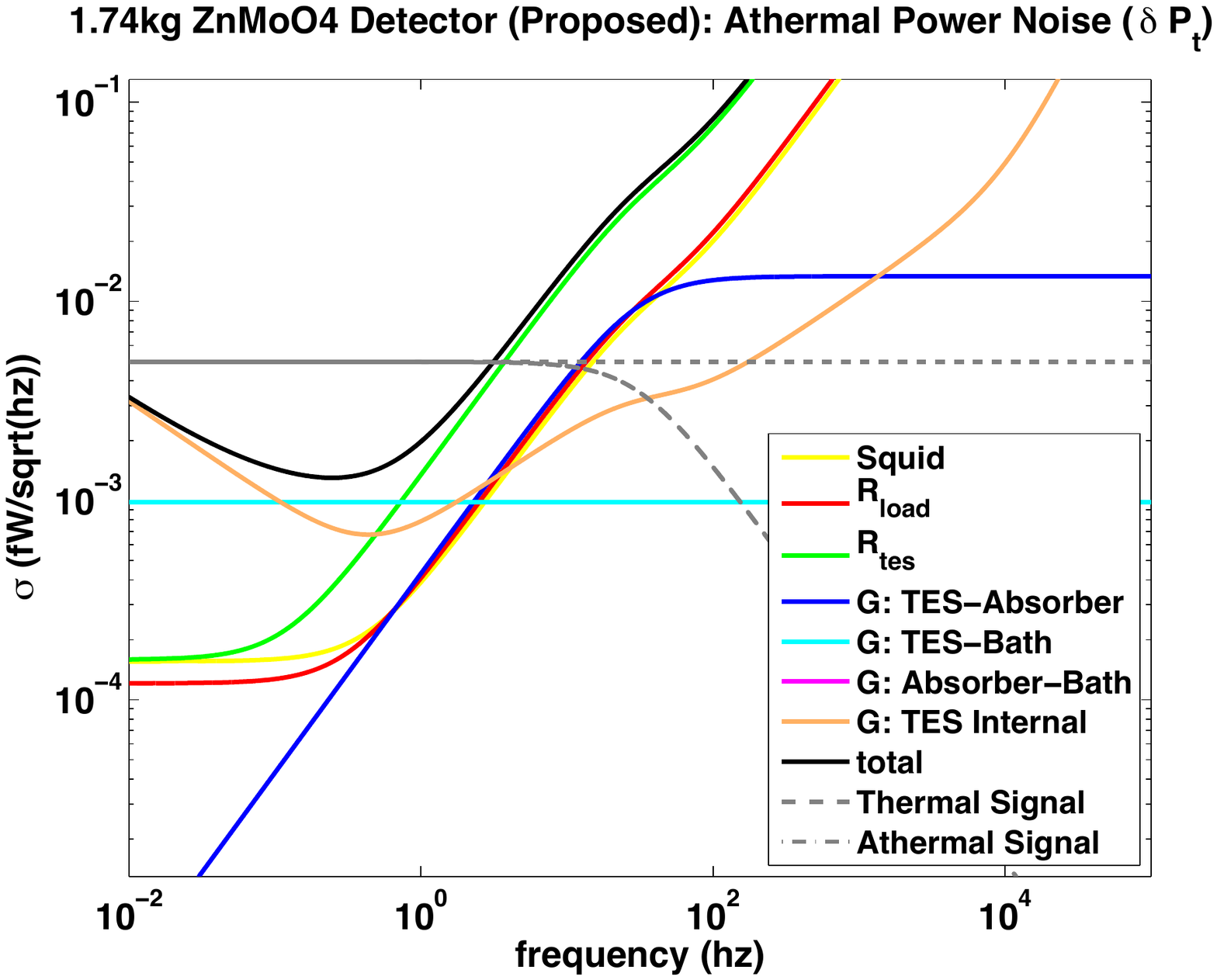}
\includegraphics[width=2.9in]{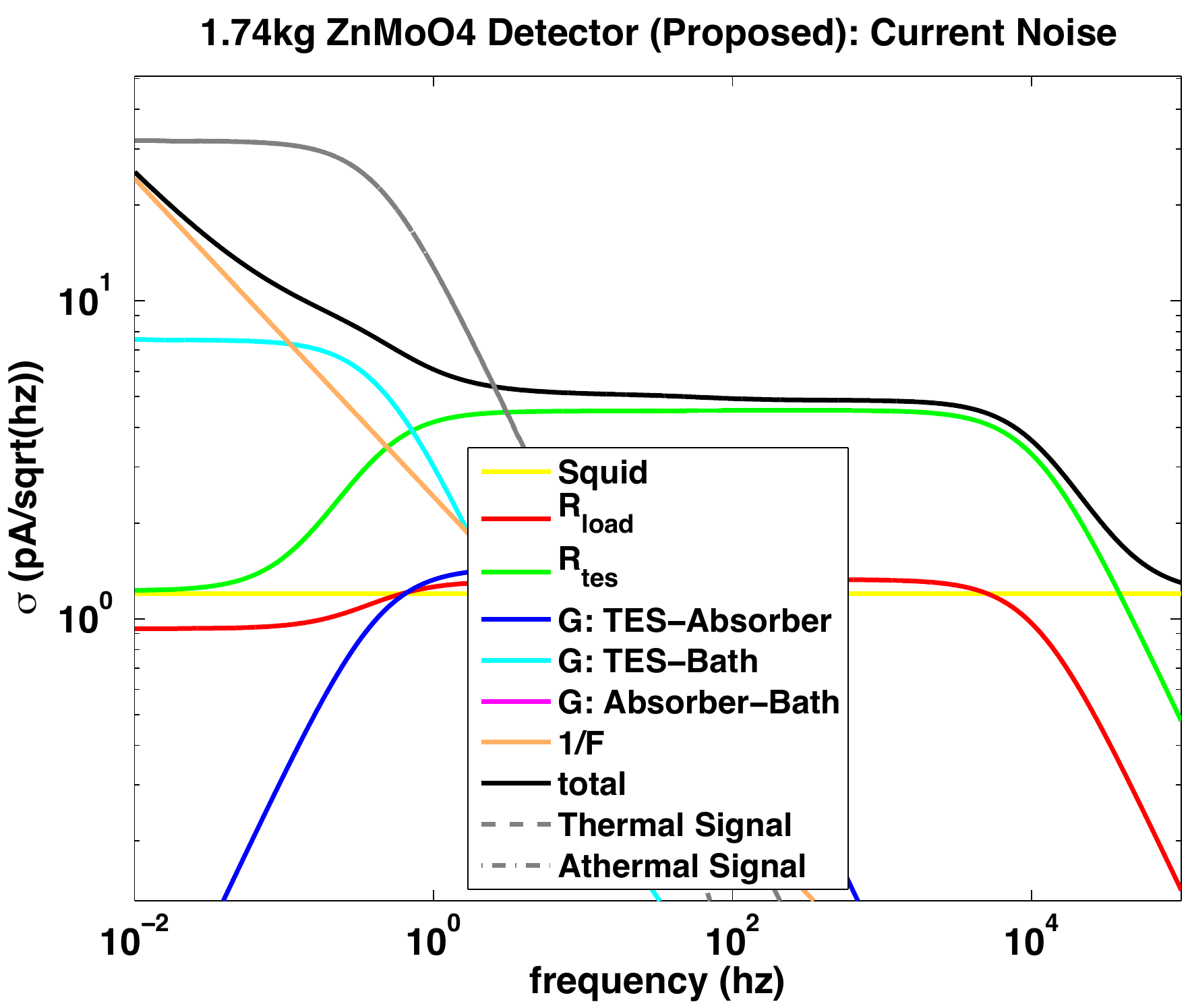}
\caption{\small{Simulated noise referenced  to TES power (top) and current (bottom) for a 1.74kg ZnMoO$_{4}$ detector.}}
\label{fig:CUORENew_PSD}
\end{figure}

\begin{figure}[h!]
\centering
\includegraphics[width=2.9in]{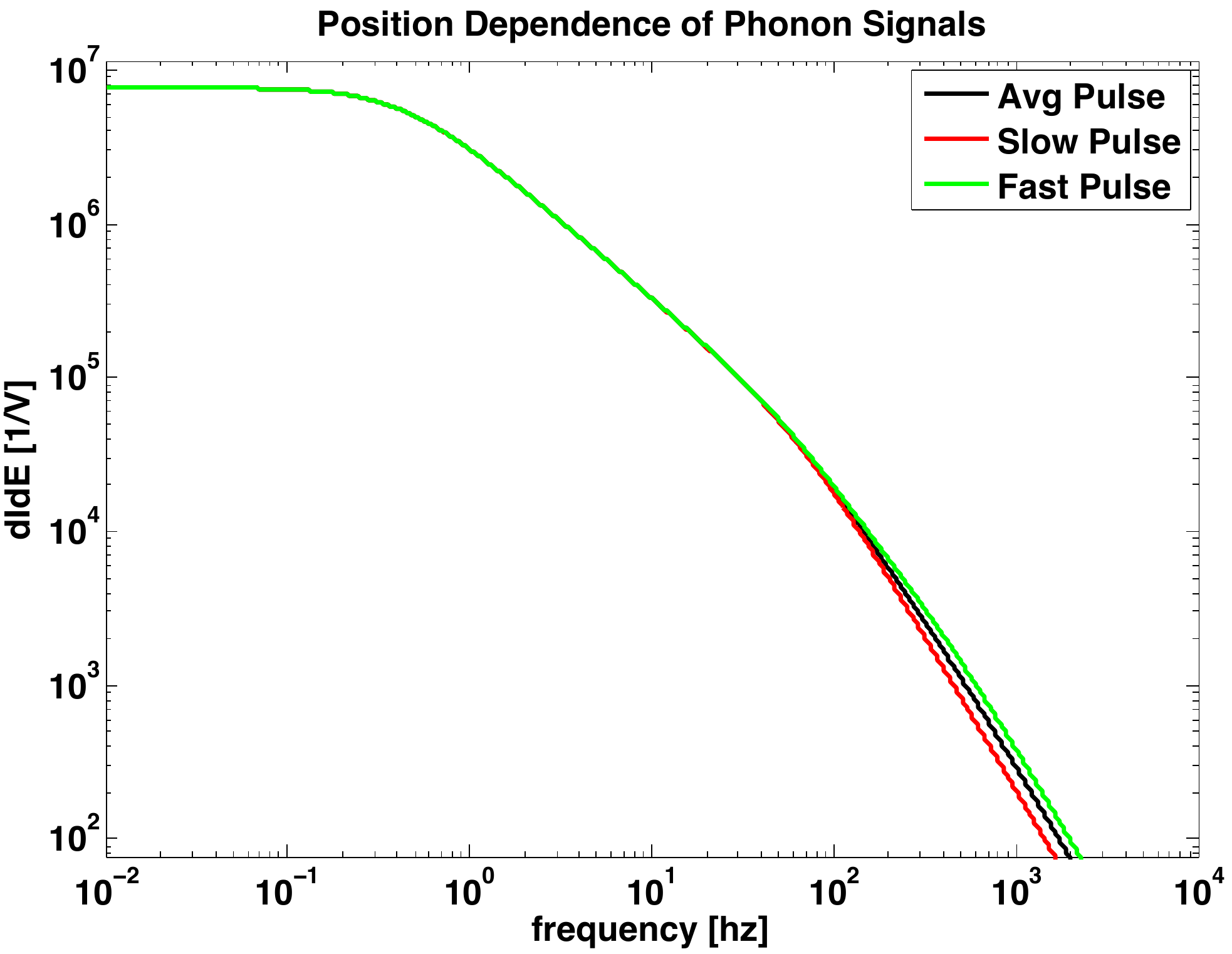}
\caption{\small{Simulated pulse shapes for events from different locations in the absorber for a 1.74 kg ZnMoO$_{4}$ detector.}}
\label{fig:xyzPulseShape}
\end{figure}

The simulated baseline energy resolution of 3.54 eV is clearly sufficient for double beta decay (Simulated Noise PSDs in Fig. \ref{fig:CUORENew_PSD}). For a rough estimate of the position sensitivity, we have  simulated position-dependent pulse shapes for $\epsilon$ = 0, 5, and 10\% (Eq.~\ref{eq:dIdExyz}, Fig. \ref{fig:xyzPulseShape}), and using the standard optimum sensitivity estimator we estimate a $\frac{\Delta E}{E}$ of 6$\times$10$^{-4}$ which is just adequate for current purity TeO$_{2}$ and ZnMoO$_{4}$.  Further, SuperCDMS has been able to suppress unwanted position dependence in their phonon energy estimators by an additional factor of 2--4 using non-stationary optimum filters or multiple template optimum filters. Techniques such as these could be even more viable for double-beta decay detectors since the residual dependence in SuperCDMS is due largely to variations in Luke phonon production due to ionized e$^{-}$ trapping \cite{MCP_Thesis}. 

This slight sensitivity to athermal phonons that systematically biases our energy estimator does have one hidden silver lining:  our devices should have significantly improved timing and pulse-shape rejection from the largely decoupled NTDs currently used by CUORE/LUCIFER, a significant advantage.  We find that position-dependent systematics dominate our start-time sensitivity and limit our start-time resolution to 170 $\mu$s ($\sigma_{to}$ = 50 $\mu$s if noise limited).

\section{Single Photon Light Detector}

These very same design rules can also be applied to (hopefully) significantly improve the performance of very low-temperature, large-area light detectors which are used to distinguish electronic recoils from nuclear recoils for both dark matter and double-beta decay detectors. For CRESST, a single-photon sensitive detector would be very beneficial because their low-mass WIMP sensitivity is suppressed by electron-recoil backgrounds leaking into the WIMP signal region \cite{CRESSTII_2014_LowMass}, while for double-beta decay experiments, single-photon sensitivity could potentially be used for electron/nuclear-recoil discrimination in non-scintillating crystals like TeO$_{2}$ via Cherenkov light collection \cite{CUORE_Cherenkov}.

The 30 mm $\times$ 30 mm Si chip used in the original CRESST light detectors has a phonon heat capacity of only 220 fJ/K at 10 mK, and consequently the parasitic heat capacity of the Au wirebond between the Si TES chip and the large Si wafer (unfortunately) totally dominates the overall sensor heat capacity as outlined in Table \ref{tab:OptLightDet}.  On the brightside, this allows us to increase the size of the Si chip to 80 mm diameter, thereby improving the  scintillation collection efficiency without significant loss of energy sensitivity.  The Au thin-film meander which thermally connects the TES to the bath was chosen so that $\tau_\mathit{eff} \sim$ 50 ms, long enough to be minimally sensitive to both the precise location of the scintillation absorption in the Si  and to stochastic fluctuations in the scintillation creation/absorption time (O(1 ms) in many crystals), yet small enough to retain significant start-time information that could be vital for application of this technology to double-beta decay with ZnMoO$_{4}$ crystals\cite{CUORE_ZnMoO4}.  Specifically, the expected noise-limited optimal-filter start-time resolution for a $\sim$ 1.5 keV scintillation signal accompanying a  3 MeV electron recoil in ZnMoO$_{4}$  is found to be 2.2 $\mu s$ (assuming instantaneous photon release/capture). Clearly, this is a drastic underestimate for slowly scintillating crystals, but it suggests that pileup rejection with the photon detector could be viable in fast scintillating crystals.

\begin{table}
\centering
\small
\begin{tabular}{| l | m{2.8cm} | m{2.4cm} | l |}
\hline
\multicolumn{4}{| c |}{Si absorber} \\
\hline														
$V_{a}$	& Volume								&														& $\pi$(4 cm)$^{2}$x525 $\mu$m \\ 
$M_{a}$	& Mass								&														& 6.0 g\\ 
$T_{ao}$	& Operation temperature 					&														& 9.3 mK \\ 
$C_{a}$	& Heat capacity 						& $\Gamma_{Si} V_{a} T_{ao}^{3}$								& 1.6 $\frac{pJ}{K}$\\  
\hline
\multicolumn{4}{| c |}{Au thin film thermal link between  TES and bath} \\
\hline
$l_{tb}$		& Length							&														& 6.0 mm  \\ 
$A_{tb}$		& Cross sectional area				&														& 5 $\mu$m x 300 nm \\
$G_{tb}$		& Thermal conductance from TES to bath	& $\frac{v_{f}d_{e}}{3} \Gamma_{Au} T_{to} \frac{A_{tb}}{l_{tb}}$		& 23 $\frac{pW}{K}$\\
\hline
\multicolumn{4}{| c |}{Au thermalization film on absorber (p1)} \\
\hline
$V_{p1}$		&  volume						& 														&$\pi$(3 mm)$^{2}$x100 nm\\
$C_{t\,p1}$	&  p1 component of TES heat capacity 	& $\gamma_{Au}V_{p1} T_{to}$								&1.9 $\frac{pJ}{K}$ \\
$G_{ta}$		& Thermal conductance between TES and absorber %
			   								& $n_{ep}\Sigma_{ep}V_{p1}T_{to}^{n_{ep}-1}$						&452 $\frac{pW}{K}$\\
\hline
\multicolumn{4}{| c |}{Au wirebond between  TES and absorber} \\
\hline
$l_{w}$		& length							&														& 1.0 cm \\ 
$A_{w}$		& Cross sectional area				&														& $\pi$(7.5 $\mu$m)$^{2}$\\
$G_{w}$		& Thermal conductance 				& $\frac{v_{f} d_{e}}{3} \Gamma_{Au} T_{to} \frac{A_{w}}{l_{w}}$  		& 5.4 $\frac{nW}{K}$\\
$C_{t\,w}$		& Bond component of TES heat capacity 	&														& 1.16 $\frac{pJ}{K}$\\
\hline
\multicolumn{4}{| c |}{W  TES} \\
\hline
$A_{t}$		& Cross sectional area				&														& 1520 $\mu$m x 150 nm \\
$l_{t}$		& Length 							&														& 300 $\mu$m \\
$T_{c}$		& Transition temperature				&														& 10 mK \\
$I_{to}$		& Operating current					& 														& 1.48 $\mu$A \\
$T_{to}$		& Operating temperature				&														& 9.3 mK \\
$G_{t\,int}$	& Internal TES conductance			& $\frac{L_{wf}}{\rho_{W}} \frac{A_{t}}{l_{t}/2} T_{to}$					& 4.9 $\frac{nW}{K}$\\
$C_{t \, W}$ 	& W component of TES heat capacity	& $f_{sc} \Gamma_{W} V_{t} T_{to}$								& 0.18 $\frac{pJ}{K}$ \\  
$R_{n}$		& Normal resistance					& $\rho_{W} \frac{l_{t}}{A_{t}}$									& 100 $m\Omega$\\
$R_{to}$ 		& Operating resistance				& $R_{n}$/4												& 20 $m\Omega$\\
$\alpha$		& Thermal sensitivity					& $\frac{T_{to}}{R_{to}} \left. \frac{\partial R}{\partial T_{t}} \right|_{I_{t o}}$	& 16.4 \\
$\beta$ 		& Current sensitivity					& $\frac{I_{to}}{R_{to}} \left. \frac{\partial R}{\partial I_{t}} \right|_{T_{t o}}$	& 0.03\\
\hline
\multicolumn{4}{| c |}{ Dynamical time constants} \\
\hline
$\tau_{L/R}$		& Squid Inductor Time Constant		&													& 15 $\mu$s \\
$\tau_\mathit{eff}$		& Sensor Fall Time					&													& 56ms\\
$\tau_{sa}$		& Absorber/TES Coupling Time Constant 	&													& 2.6ms\\				
\hline
\multicolumn{4}{| c |}{ Estimated Resolution} \\
\hline
$\sigma_{E}$		& Estimated Baseline Energy Resolution	&													& 0.36eV\\
$\sigma_{to}$ 		& Estimated Timing  Resolution @ 1.5keV &													& 2.2$\mu$s\\
\hline
\end{tabular}
\caption{Optimized Light Detector}
\label{tab:OptLightDet}
\end{table}

The expected energy-resolution performance of 0.36~eV is well below the single-photon quanta energy of $\sim$2~eV for most scintillating crystals, and consequently this device can be characterized as a highly efficient large-area single-photon detector. If additional sensitivity is required, the TES could be directly fabricated onto the Si chip for improved performance.  Of course, one would then lose the fabrication rate benefit of being able to make $\sim$ 20 devices per wafer. On the other hand, Si wafer processing is standard.

\section{Conclusions}

The small heat capacitance of insulating and semiconducting crystals at very low temperatures suggests that eV-scale energy resolution is possible in massive, kilogram thermal detectors operating around 10 mK.  Unfortunately,  with achieved sensitivities 2--3 orders of magnitude worse than these expectations, the current generation of detectors have yet to fully realize this potential. The dominant culprit for this discrepancy is that electron-phonon coupling in both TESs and NTDs drops rapidly with temperature, and thus the electron system within the sensor decouples from the phonon system within the absorber leading to the absorber phonon signal bandwidth being much smaller than the sensor bandwidth and/or shunting of the signal past the sensor through parasitic thermal conductance channels.

After a detailed study of current-generation detectors, we developed prototype designs for both an 80~mm diameter Si light detector and a 1.75~kg ZnMoO$_{4}$ massive calorimeter for which the sensor bandwidth is smaller than the signal bandwidth yet still larger than an expected $1/f$ noise threshold as well as the bandwidth requirements due to event rate. Consequently, we estimate device sensitivities of 0.4 eV and 3.4 eV that nearly match the expected scaling law performance.  Further, we have shown that these lower $G_{tb}$ designs have achievable parasitic power requirements via comparison to detectors made by SPICA/SAFARI for infrared spectrometry applications. 

Additionally, for fabrication simplicity and suppression of cosmogenic backgrounds, both of these designs use a single layer Au thin-film pad that is deposited directly onto the absorber (using simple shadow mask techniques) for electron-phonon coupling rather than fabricating the entire TES onto the absorber. This pad is then thermally connected via an Au wirebond to a separately fabricated TES chip.

Finally, we have shown that position systematics on energy estimators can still be suppressed to the level that they are subdominant compared to thermalization systematics in both TeO$_{2}$ and ZnMoO$_{4}$  crystals. This is achieved despite purposely designing our detector to have large coupling between the electronic system within the TES and the phonon system in the absorber by requiring that the sensor bandwidth be smaller than the athermal phonon collection bandwidth. 

\section{Acknowledgements}
The authors would like to thank Ray Bunker, Blas Cabrera, Raul Hennings-Yeomans,  Kent Irwin, Alexandre Juillard, Yury Kolomensky, Nader Mirabolfathi, Wolfgang Rau, and Philippe Di Stefano for interesting and insightful discussions and comments.

\appendix
\section{Pertinent Material Properties}
\label{appendix:MatProp}

\begin{table}[h!]
\centering
\small
\begin{tabular}{| l | m{4cm} | m{2.8cm}|}
\hline
\multicolumn{3}{|c|}{Tungsten Properties} \\
\hline
$\Gamma_{W}$	 			& electronic heat capacity coefficient 				& 107 $\frac{J}{m^3K^2}$ \cite{CRESST_2007_LTD7_DetPerformance} \\
$n_{ep}$ 					& electron-phonon thermal power exponent 			& 5  \cite{SHart_PSep}\\
$\Sigma_{epW}$ 				& electron-phonon coupling coefficient 				& 0.32x10$^9 \frac{W}{m^3 K^5}$ \cite{SHart_PSep} \\
$\rho_{W}$					& W electrical resistivity							&  7.6x10$^{-8} \Omega m$ \cite{CRESST_2005_FirstLimits} \\
f$_{sc}$						& superconductor/normal metal heat capacity ratio		& 2.54 \\
\hline
\multicolumn{3}{|c|}{Gold Properties} \\
\hline
$\Gamma_{Au}$				& electronic heat capacity coefficient 				& 66 $\frac{J}{m^3K^2}$ \cite{AM} \\
n$_{eAu}$					& free electron density 							& 5.90x10$^{28} \frac{1}{m^3}$ \cite{AM}\\
v$_{fAu}$					& fermi velocity									& 1.4x10$^{6} \frac{m}{s}$ \cite{AM}\\
d$_{eAu}$					& electron diffusion length (not annealed)				& $\min$(1$\mu$m,thickness) \cite{White_AuConductance}\\
$n_{ep}$					& electron-phonon thermal power exponent			& 5\\
$\Sigma_{epAu}$ 				& electron-phonon coupling coefficient 				& 3.2x10$^9 \frac{W}{m^3 K^5}$ \\
\hline
\multicolumn{3}{|c|}{Silicon Properties} \\
\hline
$\rho_{nSi}$		& atomic number density													& 5.00x10$^{28} \frac{1}{m^3}$ \\ 
$T_{DSi}$		& Debye temperature													& 645K \cite{wikipedia} \\
$\Gamma_{Si}$	& phonon heat capacity coefficient: $\frac{12 \pi^{4} k_{b} \rho_{nSi}}{5 T_{DSi}^{3}}$ 	& 0.6025 $\frac{J}{m^3K^4}$ \\
\hline
\multicolumn{3}{|c|}{CaWO$_{4}$ Properties} \\
\hline
$\rho_{\text{\tiny{n,CaWO$_{4}$}}}$			& unit cell number density																		& 1.267x10$^{28} \frac{1}{m^3}$\\ 
$T_{\text{\tiny{D,CaWO$_{4}$}}}$			& Debye temperature																		& 250K \cite{CRESST_2009}\\
$\Gamma_{\text{\tiny{CaWO$_{4}$}}}$		& phonon heat capacity coefficient: $\frac{12 \pi^{4} k_{b} \rho_{\text{\tiny{n,CaWO$_{4}$}}}}{5 T_{\text{\tiny{D,CaWO$_{4}$}}}^{3}}$	& 2.6182 $\frac{J}{m^3 K^4}$ \\
\hline
\multicolumn{3}{|c|}{ZnMoO$_{4}$ Properties} \\
\hline
$\rho_{\text{\tiny{n,ZnMoO$_{4}$}}}$			& unit cell number density																		& 1.267x10$^{28} \frac{1}{m^3}$\\ 
$T_{\text{\tiny{D,ZnMoO$_{4}$}}}$			& Debye temperature																		& 250K \cite{CRESST_2009}\\
$\Gamma_{\text{\tiny{ZnMoO$_{4}$}}}$		& phonon heat capacity coefficient: $\frac{12 \pi^{4} k_{b} \rho_{\text{\tiny{n,ZnMoO$_{4}$}}}}{5 T_{\text{\tiny{D,ZnMoO$_{4}$}}}^{3}}$	& 2.6182 $\frac{J}{m^3 K^4}$ \\
\hline
\end{tabular}
\caption{Pertinent Material Properties}
\label{tab:MatProp}
\end{table}

\bibliographystyle{aipnum4-1}
\bibliography{Bib_CRESST}

\end{document}